\documentclass[onecolumn]{IEEEtran}
\ifCLASSINFOpdf
\else
\fi
\usepackage[cmex10]{amsmath}
\usepackage{array}
\usepackage{amsmath,amssymb,amsthm,array,graphicx,verbatim}
\usepackage{latexsym,url}
\usepackage{color,array,graphicx,verbatim,xtab}
\usepackage{color}
\usepackage{setspace}
\usepackage[breaklinks,bookmarks,bookmarksnumbered,bookmarksopen,bookmarksopenlevel=2]{hyperref}

\newtheorem{theorem}{Theorem}
\newtheorem{lemma}{Lemma}
\newtheorem{definition}{Definition}

\usepackage{enumerate}
\usepackage[numbers, sort&compress]{natbib}

\newtheorem{prop}[theorem]{Property}
\newtheorem{thm}{Theorem}
\newenvironment{myproof}[1][\proofname]{\noindent\ignorespaces{\textbf{#1:} }}{}

\begin{document}
%
\title{On The Optimality of Myopic Sensing in Multi-State Channels
\textsuperscript{1}}

%
\author{Yi Ouyang\textsuperscript{2} and Demosthenis Teneketzis\textsuperscript{2}}

\maketitle

\footnotetext[1]{A preliminary version of this paper appeared in the proceedings of \textit{50th annual Allerton Conference on Communication, Control, and Computing}}
\footnotetext[2]{Y. Ouyang and D. Teneketzis are with the Department of EECS,University of Michigan, Ann Arbor, MI}

\begin{abstract}
We consider the channel sensing problem arising in opportunistic scheduling over fading channels, cognitive radio networks, and resource constrained jamming.
The communication system consists of $N$ channels. Each channel is modeled as a multi-state Markov chain (M.C.).
At each time instant a user selects one channel to sense and uses it to transmit information.
A reward depending on the state of the selected channel is obtained for each transmission.
The objective is to design a channel sensing policy that maximizes the expected total reward collected over a finite or infinite horizon.
This problem can be viewed as an instance of a restless bandit problem, for which the form of optimal policies is unknown in general.
We discover sets of conditions sufficient to guarantee the optimality of a myopic sensing policy;
we show that under one particular set of conditions the myopic policy coincides with the Gittins index rule.
\end{abstract}


%
\IEEEpeerreviewmaketitle

\section{Introduction and Literature Survey}
\subsection{Motivation}

Consider a communication system consisting of $N$ independent channels. Each channel is modeled as a $K$-state Markov chain (M.C.) with known matrix of transition probabilities.
At each time period a user selects one channel to sense and uses it to transmit information.
A reward depending on the state of the selected channel is obtained for each transmission.
The objective is to design a channel sensing policy that maximizes the expected total reward
(respectively, the expected total discounted reward) collected over a finite (respectively, infinite) time horizon.

The above channel sensing problem arises in cognitive radio networks, opportunistic scheduling over fading channels, as well as on resource-constrained jamming (\cite{zhao2007survey}).
In cognitive radio networks a secondary user may transmit over a channel only when the channel is not occupied by the primary user.
Thus, at any time instant, state $1$ of the M.C. describing the channel can indicate that the channel is occupied at $t$ by the primary user, and states $2$ through $K$ indicate the quality of the channel that is available to the secondary user at $t$.
In opportunistic transmission over fading channels, states $1$ through $K$ of the M.C. describe, at any time instant, the quality of the fading channel.
In resource-constrained jamming a jammer can only jam one channel at a time, and any given jamming/channel sensing policy results in an expected reward for the jammer due to successful jamming.

The above channel problem is also an instance of a restless bandit problem (\cite{whittle1988restless,gittins2011multi}). Restless bandit problems arise in many areas, including wired and wireless communication systems, manufacturing systems, economic systems, statistics, etc (see \cite{whittle1988restless,gittins2011multi}).

\subsection{Related Work}
The channel sensing problem has been studied in \cite{zhao2007decentralized} using a partially observable Markov decision process (POMDP) framework.
For the case of two-state channels, the myopic policy was studied in \cite{zhao2008myopic},
where its optimality was established when the number of channels is two.
For more than two channels, the optimality of the myopic policy was proved in \cite{javidi2008optimality} under certain conditions on channel parameters.
This result for the two-state channel was extended in \cite{ahmad2009optimality} using a coupling argument to establish the optimality under a relaxed ``positively correlated" condition.
In \cite{ahmad2009multi}, under the same ``positively correlated" channel condition, 
the myopic policy was proved to be optimal for two-state channels when the user can select multiple channels at each time instance. 

For general restless bandit problems, there is a rich literature; however, very little is known about the structure of optimal policies for this class of problems in general.
In \cite{whittle1988restless} it has been shown that the Gittins index rule (see \cite{gittins2011multi},\cite{gittins1979bandit} for the definition of the Gittins index rule) is not optimal for a general restless bandit problems.
Moreover, this class of problem is PSPACE-hard in general \cite{papadimitriou1994complexity}.
In \cite{whittle1988restless} Whittle introduced an index policy (referred to as Whittle's index) and an ``indexability condition"; the asymptotic optimality of the Whittle index was addressed in \cite{weber1990index}.
Issues related to Whittle's indexability condition were discussed in \cite{whittle1988restless,weber1990index,nino2007dynamic,liu2010indexability,gittins2011multi}.
For the two-state channel sensing problem, Whittle's index was computed in closed-form in \cite{liu2010indexability}, where performance simulation of that index was provided.
For some special classes of restless bandit problems,
the optimality of some index-type policies was established under certain conditions (see \cite{lott2000optimality,ehsan2009server}).
Approximation algorithms for the computation of optimal policies for a class of restless bandit problems similar to the one studied in this paper were investigated in \cite{guha2010approximation}.

A preliminary version of this paper appeared in the proceedings of the 50th Allerton conference 
on Control, Communication, and Computing (see \cite{ouyang}).

\subsection{Contribution of the Paper}
In this paper we identify sets of conditions under which the sensing policy that chooses at every time instant the best (in the sense of stochastic dominance \cite{marshall2010inequalities}) channel maximizes the total expected reward (respectively, the expected total discounted reward) collected over a finite (respectively, infinite) time horizon.
We also show that under one particular set of conditions the above-described policy coincides with the Gittins index rule, that is, the rule according to which the user selects at each time instant the channel with the highest Gittins index.
Since our model is more general than previously studied models (\cite{ahmad2009optimality}), our results are a contribution to the state of the art in cognitive radio networks, opportunistic scheduling and resource-constrained jamming.
Furthermore, the results of this paper are a contribution to the state of the art of the theory of restless bandits (see for example \cite{whittle1988restless,gittins2011multi}).
The optimization problem formulated in this paper is a restless bandit problem. Restless bandit problems are difficult to solve;
very little is known about the nature of the optimal solution of these problems (\cite{gittins2011multi}).
Our results reveal instances of restless bandit problems where: (i) the optimal allocation rule is the myopic policy; and (ii) the myopic policy is optimal and coincides with the Gittins index rule.

\subsection{Organization}
The rest of this paper is organized as follows.
In Section \ref{sec:mod}, we present the model and the formulation of the optimization problem associated with the channel sensing problem.
In Section \ref{sec:char} we discuss the salient features of the optimization problem formulated in Section \ref{sec:mod} and show that it is an instance of a restless bandit problem.
In Section \ref{sec:finite}, we consider the finite horizon problem and identify sets of conditions sufficient to guarantee the optimality of the myopic policy.
In Section \ref{sec:infinite}, we extend the results of Section \ref{sec:finite} to the infinite horizon problem.
In Section \ref{sec:twostate}, we show that the result for two-state channels in \cite{ahmad2009optimality} is a special case of the more general results presented in this paper.
In Section \ref{sec:gittins} we show that under one particular set of conditions the myopic policy coincides with the Gittins index rule.
We conclude in Section \ref{sec:conclusion}. The proofs of several intermediate results needed to establish the optimality of the myopic policy appear in the Appendices A-D.
\section{Model and Oprimization Problems} 
\label{sec:mod}
\subsection{The Model}
\label{sub:model}
Consider a communication system consisting of $N$ identical channels. Each channel is modeled as a $K$-state Markov chain (M.C.) with (the same) matrix of transition probabilities $P$,
\begin{align}
P = \left[
\begin{array}{r r r r}
p_{11}& p_{12}&\cdots &p_{1K}\\
p_{21}& p_{22}&\cdots &p_{2K}\\
\vdots & \vdots & \ddots &\vdots\\
p_{K1}& p_{K2}& \cdots &p_{KK}
\end{array}
\right]
 = 
  \left[
\begin{array}{r}
P_1\\
P_2\\
\vdots\\
P_K
\end{array}
\right],
\end{align}
where $P_1,P_2,...,P_K$ are row vectors.
The $K$ channel states model the channel's quality. For example, state $K$ may denote the highest quality state, state $1$ the lowest quality state, and states $2,3,...,K-1$ are medium quality states.
We assume that the channel's quality increases as the number of its state increases.
We want to use this communication system to transmit information. For that matter, at each time $t= 0,1,...,T$, we can select one channel, observe its state, and use it to transmit information.

Let $X^n_t$ denote the state of channel $n$ at time $t$, and 
let $U_t$ denote the decision made at time $t$; $U_t \in \{1,2,...,N\}$, where $U_t = n$ means that channel $n$ is chosen for data transmission at time $t$.

Initially, before any channel selection is made, we assume that we have probabilistic information about the state of each of the $N$ channels.
Specifically, we assume that at $t=0$ the decision-maker (the entity that decides which channel to sense at each time instant) knows the probability mass function (PMF) on the state space of each of the $N$ channels; that is, the decision-maker knows
\begin{align}
\pi_0:=(\pi^1_0,\pi^2_0,...,,\pi^N_0),
\end{align}
where
\begin{align}
&\pi^n_0 := (\pi^n_0(1),\pi^n_0(2),...,\pi^n_0(K)), n = 1,2,...,N,\\
\text{and }&\pi^n_0(i) := P(X^n_0=i), i=1,2,...,K.
\end{align}
Then, in general, 
\begin{align}
&U_0 = g_0(\pi_0) \label{g:def1}\\
&U_t = g_t(Y^{t-1},U^{t-1}),t=1,2,... 
\label{g:def2}
\end{align}
where 
\begin{align}
Y^{t-1} := &(Y_0,Y_1,...,Y_{t-1}),U^{t-1} := (U_0,U_1,...,U_{t-1}),
\end{align}
and $Y_t = X^{U_t}_t$ denotes the observation at time $t$; $Y_t$ gives the state of the channel that is chosen at time $t$ (that is, if $U_t = 2$, $Y_t$ gives the state of channel $2$ at time $t$).
\\
Let $R(t)$ denote the reward obtained by the transmission at time $t$. We assume that $R(t)$ depends on the state of the channel chosen at time $t$. That is 
\begin{equation}
R(t) = R_i, i = 1,2,...,K,
\end{equation}
if the state of the channel chosen at $t$ is $i$.
\subsection{The Optimization Problems}
Under the above assumptions, the objective is to solve:\\
(i) the finite horizon $(T)$ optimization problem (P1)
\\
\textit{Problem (P1)}\label{problemP1}
\begin{align}
\max_{g\in\mathcal{G}} E^g\{\sum_{t=0}^{T} \beta^t R(t)\};
\end{align}
and (ii) its infinite horizon counterpart, problem (P2)\\
\textit{Problem (P2)}\label{problemP2}
\begin{align}
\max_{g\in\mathcal{G}} E^g\{\sum_{t=0}^{\infty} \beta^t R(t)\},
\end{align}
\\
where $\beta$ is the discount factor ($0< \beta < 1$) and
$\mathcal{G}$ is the set of all channel sensing strategies $g$ defined by (\ref{g:def1})-(\ref{g:def2}).
\\
Problems (P1) and (P2) are centralized stochastic optimization problems with imperfect information.
Therefore, an information state for the decision-maker at time $t,t=1,2,...$ is the conditional PMF
(see \cite{kumar1986stochastic}, Chapter 6)
\begin{align}
&\pi_t :=(\pi^1_t,\pi^2_t,...,\pi^N_t),\label{eq:pi1} \\
&\pi^n_t := (\pi^n_t(1),\pi^n_t(2),...,\pi^n_t(K)), n=1,2,...,N,\label{eq:pi2}\\
&\pi^n_t(i) := P(X^n_t=i|Y^{t-1},U^{t-1}), i=1,2,...,K.\label{eq:pi3}
\end{align}
The information state $\pi_t$ evolves as follows. 
If $U_t=n, Y^{n}=i$, then
\begin{align}
&\pi^n_{t+1}=P_i,\\
&\pi^j_{t+1}=\pi^j_{t}P, 
\end{align}
for all $ j \neq n$.
From stochastic control theory \cite{kumar1986stochastic} we know that 
for problems (P1) and (P2) we can restrict attention (without any loss of optimality) to
separated policies, that is, policies of the form
\begin{align}
g :=(g_0,g_1,...),
\end{align}
where $U_t=g_t(\pi_t)$ for all $t$.\\
Consequently, problems (P1) and (P2) are equivalent to the following problems (P1') and (P2'), respectively:\\
\textit{Problem (P1')}\label{problemP1'}
\begin{align}
\max_{g\in\mathcal{G}_s} E^g\{\sum_{t=0}^{T} \beta^t R(t)\},
\end{align}
\textit{Problem (P2')}\label{problemP2'}
\begin{align}
\max_{g\in\mathcal{G}_s} E^g\{\sum_{t=0}^{\infty} \beta^t R(t)\},
\end{align}
where $\mathcal{G}_s$ is the set of separated policies.
\subsection*{Remark:}
One separated policy the performance of which we will analyse in this paper is the ``myopic policy" that we define as follows.\\
Let $\Pi$ denote the set of PMFs on the state space $S=\{1,2,...,K\}$. We define the concept of stochastic dominance/order.
Stochastic dominance $\geq_{st}$ between two row vectors $x,y \in \Pi $ is defined as follows: \\
$x \geq_{st} y$ if
\begin{align}
&\sum_{j = i}^K x(j) \geq \sum_{j = i}^K y(j)\text{ , for }i = 2,3,...,K
\end{align}
Note that stochastic order is a partial order, thus, the following facts true (see \cite{marshall2010inequalities}):
\begin{description}
\item{\textbf{Fact 1}} If $x \geq_{st} y$ and $y \geq_{st} z$ then $x \geq_{st} z$.
\item{\textbf{Fact 2}} If $x \geq_{st} y$, $z \in \Pi$ and $a \in \mathbb{R}, a\geq 0$, then $ax+z \geq_{st} ay+z$.
\end{description}
\begin{definition}
\label{def:myopic}
The myopic policy $g^m:=(g^m_0,g^m_1,...,g^m_T)$ is the policy that selects at each time instant the best(in the sense of stochastic order) channel; that is, 
\begin{align}
g^m_t(\pi_t)=i \qquad\text{if } \pi^i_t \geq_{st} \pi^j_t \quad \forall j \neq i
\end{align}
\end{definition}
\section{Characteristics of the Optimization Problems}
\label{sec:char}
The optimization problems (P1') and (P2') formulated in Section \ref{sec:mod} can be viewed as an instance of a restless bandit problem as follows:

We can view the $N$ channels as $N$ arms with their PMFs as the states of the arms. The decision maker knows perfectly the states of the $N$ arms at every time instant. 
One arm is operated (selected) at each time $t$, and an expected reward depending on the state (PMF of the channel) of the selected arm is received. If arm $n$ (channel $n$) is not selected at $t$, its PMF $\pi_t^n$ evolves according to
\begin{align}
\pi_{t+1}^n = \pi_t^n P;
\end{align}
if arm $n$ (channel $n$) is selected at $t$, its PMF evolves according to 
\begin{align}
\pi_{t+1}^n = P_{Y_t}, P(Y_t = x) = \pi_{t}^n(x	).
\end{align}
The total expected reward for problem (P1') for any sensing policy $g \in \mathcal{G}_s$ can be written as
\begin{align}
J_{\beta,T}^g := & E^g[\sum_{t=0}^{T} \beta^t R(t)]= E^g[\sum_{t=0}^{T} \beta^t \pi^{U_t}_t R].
\end{align}
The total expected reward for problem (P2') for any sensing policy $g \in \mathcal{G}_s$ can be written as
\begin{align}
J_{\beta}^g := & E^g[\sum_{t=0}^{\infty} \beta^t R(t)]= E^g[\sum_{t=0}^{\infty} \beta^t \pi^{U_t}_t R],
\end{align}
where $R := [R_1,R_2,...,R_K]^T$ is the vector of instantaneous rewards. \\
Since the selected bandit process evolves in a way that differs from the evolution of the non-selected bandit processes, this problem is not a classical multi-armed bandit problem, but a restless bandit problem.

In general, restless bandit problems are difficult to solve because forward induction 
(the solution methodology for the classical multi-armed bandit problem)
does not result in an optimal policy \cite{gittins2011multi}. 
Consequently, optimal policies may not be of the index type, and 
the form of optimal policies for general restless bandit problems is still unknown.

\section{Analysis of the Finite Horizon Problem}
\label{sec:finite}
We will prove the optimality of the myopic policy $g^m$ for Problem (P1) under certain specific assumptions on the structure of the Markov chains describing the channels, on the instantaneous rewards $R = [R_1,R_2,R_3,...,R_K]^T$
and on the initial PMFs $\pi^1_0,\pi^2_0,...,\pi^N_0$
\subsection{Key Assumptions/Conditions}
\label{sub:assum}
We make the following assumptions/conditions
\begin{enumerate}[({A}1)]
\item \label{A:order}
\begin{align}
P_K \geq_{st} P_{K-1} \geq_{st},...,\geq_{st} P_1.
\end{align}
Note that the quality of a channel state increases as its number increases. Assumption (A\ref{A:order}) ensures that the higher the quality of the channel's current state the higher is the likelihood that the next channel state will be of high quality.

\item \label{A:initial} 
Let $\Pi P$ be the set of PMFs on the channel states that can be reached through transitions according to $P$, i.e.
\begin{align}
\Pi P := \{\pi P: \pi \in \Pi\};
\end{align}
note that $\Pi P$ is the convex hull of $P_1,P_2,...,P_K$.\\
At time $0$,
\begin{align} 
&\pi^1_0 ,\pi^2_0,... ,\pi^N_0 \in \Pi P \label{A:initial:eq1}\\
\text{and }& \pi^1_0 \leq_{st} \pi^2_0 \leq_{st}... \leq_{st} \pi^N_0. \label{A:initial:eq2}
\end{align}
Assumption (A\ref{A:initial}) states that initially the channels can be ordered in terms of their quality, expressed by the PMF on $S$. Moreover, the initial PMFs of the channels are in $\Pi P$.
Such a requirement ensures that the initial PMFs on the channel states are in the same space as all subsequent PMFs.
\item \label{A:K-1}
\begin{align}
P_1P \geq_{st} P_{L-1}\\
P_KP \leq_{st} P_{L}
\end{align}
Assumption (A\ref{A:K-1}) along with (A\ref{A:initial}) ensure that, any PMF $\pi$ reachable from a non-selected channel has quality between $P_{L-1}$ and $P_L$, that is $P_L \geq_{st} \pi \geq_{st} P_{L-1}$ (see also Property \ref{P:PMForder}, Section \ref{sub:propofevol}).
Here $L$ is fixed; $L$ can be any number from $2$ to $K$.

\item \label{A:reward} 
\begin{align}
&R_i - R_{i-1} \geq \beta (P_i - P_{i-1})M \geq \beta (P_i - P_{i-1})U \geq 0 \text{ for }i \neq L
\label{A:reward:eq1}\\
&R_L - R_{L-1} \geq \beta(h- P_{L-1}R)\geq 0,\label{A:reward:eqK}
\end{align}
where $M$ is the vector given by
\begin{align}
&M := U + \beta\sum_{i\geq L} p_{Ki} PU, \label{A:reward:eq2}\\
&U_i := R_i \text{ for }i=1,2,...,L-1\label{A:reward:eq3}\\
&U_i := R_i + \beta(P_i-P_{L-1})U \text{ for }i=L,L+1,...,K, \label{A:reward:eq4}
\end{align}
and $h$ is given  by
\begin{align}
& h = \frac{P_K R - \beta\sum_{i<L}p_{Ki}P_iR}{1-\beta\sum_{i<L}p_{Ki}}.\label{A:reward:eqH}
\end{align}
Assumption (A\ref{A:reward}) states that the instantaneous rewards obtained at different states of the channel are sufficiently separated (see (\ref{A:reward:eq1})(\ref{A:reward:eqK})).
Such an assumption is essential in establishing the optimality of a myopic policy. For the myopic policy to be optimal, the expected gain incurred by choosing the current best channel (say channel $n$) versus any other channel (say channel $m$) must overcompensate future losses in performance resulting in when channel $m$ is chosen instead of channel $n$. For this to happen, the rewards obtained at different states of the channel must be sufficiently separated.
\end{enumerate}
We note that (A1)-(A4) describe sets of sets of assumptions/conditions; for every value of $L, L=2,3,...,K$, we have a distinct set of conditions.

We now compare the above conditions with those made in \cite{ouyang}. When $L=K$, the above conditions are exactly the same as those in \cite{ouyang}. In \cite{ouyang} we did not address situations where $L
\neq K$ that is, situation where the quality of the information state resulting form a non-selected channel is between $P_L$ and $P_{L-1}$ for $L\neq K$. Consequently, the result of this paper subsume the results obtained in \cite{ouyang}.

Before we proceed with the analysis of Problem (P1) based on conditions (A\ref{A:order})-(A\ref{A:reward}), we show that (A\ref{A:order})-(A\ref{A:reward}) can be simultaneously satisfied. Consider the following situation:
\begin{align}
&K = 5, L = 5, N = 6, \beta =1\\
&P = 
  \left[
\begin{array}{r}
P_1\\
P_2\\
\vdots\\
P_5
\end{array}
\right] =
\left[\begin{array}{lllll}
    0.0656&    0.0458&    0.1044&    0.4745&    0.3096\\
    0.0655&    0.0458&    0.1030&    0.4454&    0.3403\\
    0.0652&    0.0457&    0.0966&    0.4019&    0.3907\\
    0.0434&    0.0336&    0.1126&    0.4102&   0.4001\\
    0.0206&    0.0205&    0.0142&    0.4475&    0.4972
\end{array}\right],\\
\end{align}
with
\begin{align}
&R = \left[\begin{array}{lllll}
    0&    1&    2&    3&    4
\end{array}\right]^T\\
&\pi^1_0 = \pi^2_0 =P_1,\pi^3_0 =P_2,\pi^4_0 =P_3,\pi^5_0 =P_4,\pi^6_0 = P_5
\end{align}
By their definition, $P_1,P_2,...,P_5$ satisfy (A\ref{A:order}).
By the definition of $\pi^1_0,\pi^2_0,...,\pi^6_0$ and the definition of $\Pi P$, (A\ref{A:initial}) is satisfied.\\
By direct computation we can show that 
\begin{align}
P_1P =& \left[\begin{array}{lllll}
0.0411&    0.0322&    0.0795&    0.4267&    0.4205
   \end{array}\right]\\
\geq_{st}& \left[\begin{array}{lllll}
    0.0434&    0.0336&    0.1126&    0.4102&    0.4001
\end{array}\right] = P_{4},
\end{align}
Moreover, $P_5P=p_{51}P_1+p_{52}P_2++...+p_{55}P_5 \leq_{st} P_5$.
Therefore, (A\ref{A:K-1}) is satisfied. \\
By direct computation, we get
\begin{align}
U = & \left[\begin{array}{lllll}
0&    1&    2&    3&    4.3214
\end{array}\right]^T\\
M = & \left[\begin{array}{lllll}
1.4997&    2.5206&    3.5577&    4.6003&    6.0815
\end{array}\right]^T\\
h = &3.7776,
\end{align}
So we can compute 
\begin{align}
&\beta(P_2-P_1)M = 0.0470 \leq R_2-R_1 \\
&\beta(P_3-P_2)M = 0.0829 \leq R_3-R_2 \\
&\beta(P_4-P_3)M = 0.0897 \leq R_4-R_3 \\
&\beta(h- P_{4}R) =   0.7766  \leq R_5-R_4
\end{align}
Therefore, (A\ref{A:reward}) is satisfied. \\
Assumptions (A\ref{A:order})-(A\ref{A:reward}) are also satisfied when
$R,P,\pi^1_0,\pi^2_0,...,\pi^6_0$, chosen as above, are slightly perturbed. It is also possible to find other ranges of values of 
$R,P,\pi^1_0,\pi^2_0,...,\pi^6_0$ which satisfy (A\ref{A:order})-(A\ref{A:reward}).
\\
Based on the above assumptions, we proceed to establish the optimality of the myopic policy $g^m$ as follows. 
In sections \ref{sub:propofevol}-\ref{sub:propoffun} we develop some preliminary results needed for our purposes. Specifically:
In section \ref{sub:propofevol} we present
three properties of the evolution of the PMFs on the channel states. 
In section \ref{sub:propofreward} we present a property of the instantaneous expected reward.
In section \ref{sub:propoffun} we define a class of ordering-based channel sensing policies $\mathcal{G}^O$ which includes the myopic policy $g^m$; using the results of sections 
\ref{sub:propofevol} and \ref{sub:propofreward} we discover four properties of the expected reward resulting from any policy in $\mathcal{G}^O$.
In section \ref{sub:myopic} we use the results of section \ref{sub:propoffun} to establish the optimality of a myopic policy for Problem (P1').
We note that all the properties developed in sections \ref{sub:propofevol} through \ref{sub:propoffun} are needed to establish the optimality of the myopic policy. 
We discuss how these properties are used to prove the optimality of the myopic policy in Section \ref{sub:diss}, after we prove the main result of this paper. The proofs of properties 1-9 appear in Appendices A-D.

\subsection{Properties of the Channels' Evolution}
\label{sub:propofevol}
Under assumptions/conditions (A1)-(A4) stated in section \ref{sub:assum}, the following properties hold.
\begin{prop} \label{P:preserveorder} 
Let $x,y \in \Pi$. Under Assumption (A\ref{A:order}),
\begin{align}
x \geq_{st} y \Longrightarrow xP \geq_{st} yP
\end{align}
\end{prop}

An implication of Property \ref{P:preserveorder} is the following. 
If at any time $t$ the information states of two channels (expressed by the PMFs on their state space) are stochastically ordered and none of these channels is sensed at $t$, then the same stochastic order between the information states at time $t+1$ is maintained.
\begin{prop} \label{P:PMForder} 
Let $\pi=xP^2 \in \Pi P^2$, $\Pi P^2:=\{\pi=xP^2, x \in \Pi \}$.
Under (A1)-(A3),
\begin{align} 
P_L \geq_{st} xP^2 \geq_{st} P_{L-1}
\end{align}
\end{prop}
Property \ref{P:PMForder} says the following. 
By condition (A2) a channel's information state (the PMF on its state space) is always in $\Pi P$.
If the channel is not sensed at time $t$, then at time $t+1$ its information state is in $\Pi P^2$,
moreover it is stochastically always between $P_{L-1}$ and $P_{L}$. If the channel is sensed at time $t$ and its observed state is larger than or equal to $L$ (respectively smaller than $L$), then at time $t+1$ this channel is in the stochastically largest (respectively stochastically smallest) information state among all channels.
\begin{prop} \label{P:canbeorder} Under (A1)-(A3), 
		we have either $\pi^n_t \leq_{st} \pi^m_t$ or $\pi^m_t \leq_{st} \pi^n_t$
		for all $n,m \in \{1,2,...,N\}$ for all $t$.
\end{prop}
Property \ref{P:canbeorder} states that under (A1)-(A3) the information states of all channels can be ordered stochastically at all times.

The proofs of Properties 1-3 appear in Appendix \ref{app:pmf}.
\subsection{A Property of the Instantaneous Expected Reward}
\label{sub:propofreward}
A direct consequence of Assumption (A\ref{A:reward}) is the following Properties of the instantaneous expected reward:
\begin{prop} \label{P:instreward} 
Let $x,y \in \Pi$. Let $v$ be a column vector in increasing order, i.e. $v_i \geq v_{i-1}$ for $i=2,3,...,K$.
If $x \geq_{st} y$, we have
\begin{enumerate}[(i)]
\item $(x-y)v \geq 0$.
\item $(x-y)M \geq (x-y)U \geq (x-y)R \geq 0$, where $M,U,R$ are defined by eqs (\ref{A:reward:eq1})-(\ref{A:reward:eq4}).
\item $(x-y)M \geq \beta(x-y)PM $.
\item If $x(i)=y(i)$ for all $i \geq L$ or $x(i)=y(i)$ for all $i < L$, we have
\begin{align}
(x-y)R \geq \beta(x-y)PM \geq \beta(x-y)PU.
\end{align}
\end{enumerate}
\end{prop}	
Part (i) of Property \ref{P:instreward} says the following. 
Consider a reward vector such that the reward increases as the quality of the channel state increases. Then the expected reward increases as the information state of the channel increases stochastically.

Part (ii) is a restatement of part (i) when the reward vector $v$ takes the values
$M-U,U-R,R$.

Part (iii) can be interpreted as follows. Consider the reward vector $M$
defined by (\ref{A:reward:eq2}).
Consider two channels, channel $i$ and channel $j$, that have information states $x$ and $y$ respectively,
such that $x \geq_{st} y$. Consider the following scenarios: 
(S1) Sense channel $i$ first, then sense channel $j$; (S2) Sense channel $j$ first, then sense channel $i$.
Then part (iii) of Property \ref{P:instreward} asserts that scenario (S1) is better than scenario (S2), that is, it is better to sense the best (in the sense of stochastic order) channel first.

Part (iv) has an interpretation similar to that of part (iii). Consider any time $t$ and two channels 
$i$ and $j$ whcih have information states $x$ and $y$, respectively, such that $x\geq_{st} y$ and $x,y$ satisfy the condition of part (iv). Assume that the reward vector at $t$ is $R$ and the reward vector at $t+1$ is $M$ such that $M_i-R_i$ is increasing in $i$. Consider scenarios (S1) and (S2) described above. 
Then part (iv) asserts that the expected reward obtained under scenario (S1) is higher than the expected reward obtained under scenario (S2); that is, it is better to sense the best (in the sense of stochastic order) channel first.
Note that Property \ref{P:instreward} refers to the situation where we have only two options, described by scenarios (S1) and (S2). Thus, the results of Property \ref{P:instreward} do not imply the optimality of the myopic policy, as in Problems (P1) we have more that two options at each time instant.

The proof of Property \ref{P:instreward} appears in Appendix \ref{app:inst}.
\subsection{Properties of the Reward Associated with Ordering-based Channel Sensing Polices}
\label{sub:propoffun}
In this section we introduce ordering-based policies and study their Properties. 
The reason for considering this class of policies is because under conditions (A1)-(A4) we obtain the following:
(i) The performance of any sensing policy can be upper-bounded by an appropriately chosen ordering-based policy (see Section \ref{sub:myopic}); thus, for the solution of the original optimization problem (Problem (P1)) we can restrict attention to ordering-based policies.
(ii) The myopic policy is an optimal ordering-based policy.
Combining (i) and (ii) we establish the optimality of the myopic policy for Problem (P1).

We note that Properties 1-4, developed so far, are essential for the discovery of the properties of ordering-based policies that lead eventually to the solution of Problem (P1) (see discussion in Section \ref{sub:diss}).

Let $\mathcal{O}$ be the set of all orderings/permutations of the $N$ channels $\{1,2,...,N\}$.
Consider the ordering-based selection function
$\hat{g} : \mathcal{O} \mapsto \{1,2,...,N\}$ and the ordering update mapping
$\hat{m} : \mathcal{O}\times \{1,2,...,K\} \mapsto \mathcal{O}$ defined as follows.\\
For every $O := (O(1),O(2),...,O(N)) \in \mathcal{O}$,
\begin{align}
&\hat{g}(O) = O(N),\\
&\hat{m}(O,y) =
\left\lbrace 
\begin{array}{ll}
O & \text{ if } y\geq L\\
SO & \text{ if } y< L
\end{array}
\right. ,
\end{align}
where $S$ is the cyclic shift operator on $\mathcal{O}$ such that
\begin{align}
S O =: (O(N),O(1),O(2),...,O(N-1))
\end{align}
Given a channel ordering $O_t\in\mathcal{O}$ at time $t$,  we define an ordering-based channel sensing policy $g^{O_t}_{t:T}:=(g^{O_t}_t,g^{O_t}_{t+1},...,g^{O_t}_{T})$ as follows.
\begin{align}
U_t= &g^{O_t}_t(O_t) = \hat{g}(O_t)=O(N)\\
O_s = & \hat{m}(O_{s-1},Y_{s-1}) \text{, for } s = t+1,t+2,...,T\\
U_s = & g^{O_t}_s(Y_{t:s-1},U_{t:s-1}) =g^{O_t}_s(O_s)= \hat{g}(O_s) \text{, for } s = t+1,t+2,...,T
\end{align}
At time $s, t\leq s \leq T$, $g^{O_t}_s$ chooses the last channel in $O_s$; the ordering $O_s$ is shifted to the right by the update mapping $\hat{m}$ whenever the observed state is less than $L$, and remains the same otherwise. As a result of the above specification of $g^{O_t}_{t:T}$, if at time $t$ channel $n$ is on the right of channel $m$ in the ordering $O_t$, channel $n$ will be sensed by policy $g^{O_t}_{t:T}$ before channel $m$.\\
Note that, the policy $g^{O_t}_{t:T}$ is not a separated policy in general. However, if the ordering $O_0=(O_0(1),O_0(2),...,O_0(N)) $ at time $0$ is such that $\pi^{O_0(1)}_0 \leq_{st}\pi^{O_0(2)}_0\leq_{st}...\leq_{st}\pi^{O_0(N)}_0 $, then $g^{O_0}_{0:T}$ is the myopic policy $g^m$, therefore; $g^{O_0}_{0:T}=g^m \in \mathcal{G}_s$, as the following Property shows.
\begin{prop} \label{P:myopic} 
At time $t=0$ consider the ordering $O_0$ such that
$\pi^{O_0(1)}_0\leq_{st}\pi^{O_0(2)}_0\leq_{st}...\leq_{st}\pi^{O_0(N)}_0$.
Then, the ordering based policy $g^{O_0}_{0:T}$  is just the myopic policy $g^m$.
\end{prop}
The validity of Property \ref{P:myopic} crucially depends on Properties \ref{P:preserveorder} and \ref{P:PMForder}, which say that stochastic order is maintained under the evolution of unobserved channels (Property \ref{P:preserveorder}), and the observed channel is either the stochastically best or the stochastically worst among all channels (Property \ref{P:PMForder}). Without Properties \ref{P:preserveorder} and \ref{P:PMForder} the myopic policy is not an ordering-based policy.\\
The proof of Property \ref{P:myopic} appears in Appendix \ref{app:fun}.

Define by $V_t(O_t,\pi^1_t,\pi^2_t,...,\pi^N_t)$ to be the expected reward collected from time $t$ up to and including $T$ due to the ordering-based policy $g^{O_t}_{t:T}$. That is,
\begin{align}
V_t(O_t,\pi^1_t,\pi^2_t,...,\pi^N_t)
:= E^{g^{O_t}_{t:T}}[\sum_{l=t}^T \beta^{l-t}R(l)
| \pi^1_t,\pi^2_t,...,\pi^N_t  ] \label{eqVtdef}
\end{align}
Then, $V_t(O_t,\pi^1_t,\pi^2_t,...,\pi^N_t)$ can be written recursively as follows.
\begin{align}
V_T(O_t,\pi^1_T,\pi^2_T,...,\pi^N_T) = & \pi^{O_t(N)}_TR \label{fun:eq2},\\
V_t(O_t,\pi^1_t,\pi^2_t,...,\pi^N_t) = & \pi^{O_t(N)}_tR  + 
\beta\sum_{i <L}\pi^{O_t(N)}_t(i)V_{t+1}(SO_t,\pi^1_{t+1},...,\pi^N_{t+1})\nonumber\\
&\quad\quad+ \beta\sum_{i \geq L}\pi^{O_t(N)}_t(i)V_{t+1}(O_t,\pi^1_{t+1},...,\pi^N_{t+1}), \label{eqVt}\\
\text{where }\pi^n_{t+1} =&\left\lbrace
\begin{array}{ll}
P_i & \text{ for }n = O_t(N)\\
\pi^n_t P & \text{ otherwise}
\end{array}\right. .
\end{align}
The function $V_t(O_t,\pi^1_t,\pi^2_t,...,\pi^N_t)$ defined above possesses 
properties \ref{P:funrotate}-\ref{P:funlift} below. 
The proof of these Properties appear in Appendix \ref{app:fun}.
We will explain the role of these Properties in Section \ref{sub:diss} after we prove the main result on the optimality of the myopic policy in Section \ref{sub:myopic}.
\begin{prop} \label{P:funrotate}  
Let $\hat{\pi}^1_t,\pi^1_t,\pi^2_t,...,\pi^{N}_t \in \Pi P$  and $O_t \in \mathcal{O}$.\\
Define
\begin{align}
L_t(O_t,\hat{\pi}^1_t,\pi^1_t,\pi^2_t,...,\pi^N_t)
:=V_t(O_t,\hat{\pi}^1_t,\pi^2_t,...,\pi^N_t)-V_t(O_t,\pi^1_t,\pi^2_t,...,\pi^N_t)
\label{eq:L}
\end{align}
If $\hat{\pi}^1_t\geq_{st}\pi^1_t$, and $O_t(n)=1$, then for all $m< n$
\begin{align}
0 \leq
L_t(O_t,\hat{\pi}^1_t,\pi^1_t,\pi^2_t,...,\pi^N_t) -L_t(S^{-m}O_t,\hat{\pi}^1_t,\pi^1_t,\pi^2_t,...,\pi^N_t) 
\leq (\hat{\pi}^1_t-\pi^1_t)U,
\end{align} 
where $S^{-m}O_t$ is the counter-clockwise cyclic shift of $O_t$ by $m$ positions, that is,
\begin{align}
S^{-m}O_t = (O_t(m+1),O_t(m+2),...,O_t(N),O_t(1),...,O_t(m))
\end{align}
\end{prop}

\begin{prop} \label{P:funswitch}  
For $O_t\in \mathcal{O}$, define the operator $W_{nm}$ as follows.
\begin{align}
W_{nm}O_t(i) :=\left\lbrace
\begin{array}{ll}
O_t(n) & \text{ for }i=m\\
O_t(m) & \text{ for }i=n\\
O_t(i) & \text{otherwise}
\end{array}
\right. .
\label{eq:WOt}
\end{align}
If $\hat{\pi}^1_t\geq_{st}\pi^1_t$, and $O_t(n)=1$, then for $ m < n$
\begin{align}
0\leq
L_t(O_t,\hat{\pi}^1_t,\pi^1_t,\pi^2_t,...,\pi^N_t) -
L_t(W_{nm}O_t,\hat{\pi}^1_t,\pi^1_t,\pi^2_t,...,\pi^N_t)
\leq (\hat{\pi}^1_t-\pi^1_t)M
\end{align} 

\end{prop}
The meaning of Properties \ref{P:funrotate} and \ref{P:funswitch} is the following.
Restrict attention to ordering-based policies. Take any channel, say channel $1$.
Replace it with a better quality (in the sense of stochastic order) channel.
Such a replacement will result in an improvement in performance. This improvement is different for different channel orderings.
The earlier channel $1$ is used (that is, the closer to the right-most position in the ordering channel $1$ is) the higher is the improvement.
Properties \ref{P:funrotate} and \ref{P:funswitch} also provide bounds on the difference between maximum and minimum improvement.
These bounds are useful in proving Properties \ref{P:funrotate} and \ref{P:funswitch} by induction.


\begin{prop} \label{P:funswitch2}
If $\pi^{O_t(n)}_t\geq_{st}\pi^{O_t(m)}_t$, then for $ m < n$
then
\begin{align}
V_t(O_t,\pi^1_t,\pi^2_t,...,\pi^N_t) \geq
V_t(W_{nm}O_t,\pi^1_t,\pi^2_t,...,\pi^N_t)
\end{align} 

\end{prop}
Property \ref{P:funswitch2} states that if the position of two channels in any arbitrary but fixed channel ordering are interchanged so that the better (in the stochastic order sense) channel comes closer to the right-most position (i.e. it is used earlier) in the new ordering, the performance due to the ordering-based policy improves.
\begin{prop} \label{P:funlift}
For $O_t\in \mathcal{O}$, define the operator $A_{nm}$ as follows.
\begin{align}
A_{nm} O_t(i) :=\left\lbrace
\begin{array}{ll}
O_t(n) & \text{ for }i=m\\
O_t(i-1) & \text{ for }i=m+1,m+2,...,n\\
O_t(i) & \text{otherwise}
\end{array}
\right. .
\label{eq:AOt}
\end{align}
If $\pi^1_t \leq_{st} \pi^1_t P$, and $O_t(n)=1$, then
\begin{align}
V_t(A_{nm} O_t,\pi^1_t,\pi^2_t,...,\pi^N_t) -
V_t(O_t,\pi^1_t,\pi^2_t,...,\pi^N_t) \leq h-\pi^1_tP^{N-n} R
\end{align}
\end{prop}
Property \ref{P:funlift} states the following.
Suppose that a channel, say channel $1$,  is such that as long as it is not sensed its quality is continuously improving (i.e. its PMF is continuously increasing stochastically). Then, no matter how late this channel is sensed (that is, no matter how much we move the channel to the left from its initial position in the original channel ordering) the change in performance due to an ordering-based policy can not exceed a certain bound.
\subsection{Optimality of a Myopic Policy}
\label{sub:myopic}
The main result of this paper is summarized by the following theorem
\begin{thm} \label{thm:myopic}
Under assumptions (A\ref{A:order})-(A\ref{A:reward}), the myopic policy $g^m$, that is, the policy that picks at every time instant the best (in the sense of stochastic order) channel is optimal for Problem (P1).
\end{thm}
\begin{myproof} \label{pfthm:myopic}
We proceed by induction. \\
At $T$, the expected reward is the instantaneous expected reward. Since by part (ii) of Property \ref{P:instreward} a better channel (in the sense of stochastic order) gives larger instantaneous expected reward, the myopic policy $g^m$ is optimal at $T$. This establishes the basis of induction.
\\\\
The induction hypothesis is that the myopic policy $g^m$ is optimal at $t+1,t+1,...,T$. To complete the induction we need to prove that $g^m$ is optimal at $t$ (induction step).
\\\\
Without loss of generality, we assume $\pi^1_t \leq_{st} \pi^2_t\leq_{st}...\leq_{st}\pi^N_t$.
\\
Consider any policy $g$. If $g$ picks channel $n$ at time $t$, then the expected reward collected from $t$ on due to the policy $g$ is given by 
\begin{align}
&E^{g}[\sum_{l=t}^T \beta^{l-t}R(l)| \pi^1_t,\pi^2_t,...,\pi^N_t  ]  
=\pi^nR +\sum_{i=1}^K\pi^n_t(i)E^{g}[\sum_{l=t+1}^T \beta^{l-t}R(l)| 
\pi^n_{t+1} = P_i,\pi^m_{t+1}=\pi^m_tP \text{ for }m\neq n  ]. \label{pfthm:myopic:eq1}
\end{align}
By the induction hypothesis we have 
\begin{align}
&E^{g}[\sum_{l=t+1}^T R(l)
| \pi^n_{t+1} = P_i,\pi^m_{t+1}=\pi^m_tP \text{ for }m\neq n   ] \nonumber\\
\leq 
&E^{g^m}[\sum_{l=t+1}^T R(l)
| \pi^n_{t+1} = P_i,\pi^m_{t+1}=\pi^m_tP \text{ for }m\neq n  ]. \label{pfthm:myopic:eq2}
\end{align}
Using (\ref{pfthm:myopic:eq2}) in (\ref{pfthm:myopic:eq1}) we get
\begin{align}
&E^{g}[\sum_{l=t}^T \beta^{l-t}R(l)| \pi^1_t,\pi^2_t,...,\pi^N_t  ]  \nonumber\\
\leq&\pi^n_tR +\sum_{i=1}^K\pi^n_t(i)E^{g^m}[\sum_{l=t+1}^T \beta^{l-t}R(l)
| \pi^n_{t+1} = P_i,\pi^m_{t+1}=\pi^m_tP \text{ for }m\neq n  ]\nonumber\\
=& \pi^n_tR+\beta\sum_{i <L}\pi^n_t(i)V_{t+1}(SO_t,\pi^1_{t+1},...,\pi^N_{t+1}) 
+\beta\sum_{i \geq L}\pi^n_t(i)V_{t+1}(O_t,\pi^1_{t+1},...,\pi^N_{t+1}) \nonumber\\
= & V_t(O_t,\pi^1_t, ...,\pi^N_t),\label{pfthm:myopic:eq3}
\end{align}
where 
\begin{align}
&O_t = (1,2,...,n-1,n+1,...,N,n),\\
&SO_t = (n,1,2,...,n-1,n+1,...,N).
\end{align}
The inequality in (\ref{pfthm:myopic:eq3}) follows by (\ref{pfthm:myopic:eq2});
the first equality in (\ref{pfthm:myopic:eq3}) is true because of Property \ref{P:myopic}, for $s = t+1,t+2,...,T$,
$g^m_s = g^{SO_t}_s$ when $\pi^n_{t+1}=P_i,i<L $ and $g^m_s = g^{O_t}_s$ when $\pi^n_{t+1}=P_i,i\geq L $;
the last equality follows from equation (\ref{eqVt}) for $V_{t}$.
\\
Since $\pi^n_t\leq_{st} \pi^m_t$ for all $m = n+1,n+2,...,N$, repeatedly applying Property \ref{P:funswitch2} we get
\begin{align}
V_t(O_t,\pi^1_t, ...,\pi^N_t)
\leq & V_t((1,2,...,n-1,n+1,...,N-1,n,N),\pi^1_t, ...,\pi^N_t)\nonumber\\
&\vdots\nonumber\\
\leq &V_t((1,2,...,n-1,n+1,n,n+2,...,N),\pi^1_t, ...,\pi^N_t)\nonumber\\
\leq &V_t((1,2,...,n-1,n,n+1,...,N),\pi^1_t, ...,\pi^N_t)\nonumber\\
= &E^{g^m}[\sum_{l=t}^T R(l)| \pi^1_t,\pi^2_t,...,\pi^N_t  ]  \label{pfthm:myopic:eq4}
\end{align}
Combing (\ref{pfthm:myopic:eq3}) (\ref{pfthm:myopic:eq4}) we obtain
\begin{align}
&E^{g}[\sum_{l=t}^T \beta^{l-t}R(l)| \pi^1_t,\pi^2_t,...,\pi^N_t  ] \leq 
E^{g^m}[\sum_{l=t}^T \beta^{l-t}R(l)| \pi^1_t,\pi^2_t,...,\pi^N_t  ],  
\end{align}
which completes the proof.
\end{myproof}
\subsection{Discussion}
\label{sub:diss}
The key steps in establishing the optimality of the myopic policy, under the assumptions made in the problem formulation, are the following:
\begin{enumerate}[(K1)]
\item The assertion that the performance of any separated policy can be upper-bounded by the performance of an ordering-based policy. Consequently, for the solution of the original optimization problem, one can restrict attention to ordering-based policies.
\item The assertion that the performance of an ordering-based policy improves when a better (in the sense of stochastic order) channel is used earlier. This assertion implies the optimality of the myopic policy.
\end{enumerate}
The assertion of (K1) is established in Theorem \ref{thm:myopic} (its induction step).
The assertion of (K2) is established by Property \ref{P:funswitch2}, provided that the myopic policy is an ordering-based policy, and that stochastic order is maintained among all channels at every time.
The fact that the myopic policy is an ordering-based policy is ensured by Property \ref{P:myopic}.
The existence of a stochastic ordering among all channels at any time $t$ is ensured by Property \ref{P:canbeorder}.
To establish these properties we need Properties 1-9.

We now elaborate on the interdependence of Properties 1-9. 
Property \ref{P:canbeorder}, which asserts that channels can be ordered stochastically, is a consequence of Properties \ref{P:preserveorder} and \ref{P:PMForder} for the unobserved channels and the observed channel, respectively.
Properties \ref{P:preserveorder} and \ref{P:PMForder} also ensure that the myopic policy $g^m$ belongs to the class of ordering-based policies (Property \ref{P:myopic}).
Property \ref{P:funswitch2} is a special case of Property \ref{P:funswitch} when $\hat{\pi}^1_t=\pi^{O_t(m)}_t\geq_{st}\pi^1_t=\pi^{O_t(n)}_t$.
Property \ref{P:funswitch} is coupled with Properties \ref{P:funrotate} and \ref{P:funlift}, that is, 
Properties \ref{P:funrotate}, \ref{P:funswitch} and \ref{P:funlift} need to be proven simultaneously.
The proof of Properties \ref{P:funrotate}, \ref{P:funswitch} and \ref{P:funlift} requires Property \ref{P:instreward}.

The upper bounds that appear in Properties \ref{P:funrotate}, \ref{P:funswitch} and \ref{P:funlift} are essential in establishing the optimality of the myopic policy. These bounds along with condition (A\ref{A:reward}) ensure that the instantaneous advantage in expected reward obtained by the use of the myopic policy $g^m$ over any other policy $g$, overcompensates any future possible expected reward losses of $g^m$ as compared to $g$.

\section{The Infinite Horizon Problem}
\label{sec:infinite}
For the infinite horizon Problem (P2) we have the following theorem.
\begin{thm} \label{thm:myopic2}
Under assumptions (A\ref{A:order})-(A\ref{A:reward}), the myopic policy $g^m$ is optimal for Problem (P2).
\end{thm}
\begin{myproof}
From the theory of stochastic control \cite{kumar1986stochastic} we know that for Problem (P2) 
there exists a separated stationary policy $g^*$ that maximizes the total expected discounted reward. \\
Let $\pi := (\pi^1,\pi^2,...,\pi^N)$; for any stationary separated policy $g$ let 
\begin{align}
J_\beta^{g}(\pi) := E^{g}\{\sum_{t=0}^{\infty} \beta^t R(t)|\pi_0=\pi\}.
\end{align}
Then the dynamic program for Problem (P2) is
\begin{align}
J_\beta^{g^*}(\pi) = \max_{n=1,2,...,N}\left\lbrace \pi^n R 
                  + \beta E \{J_\beta^{g^*}(\pi_1) | \pi_0 = \pi, U_0=n\} \right\rbrace,
                  \label{infh:eq1}
\end{align}
where $\pi_0,\pi_1$ are defined by (\ref{eq:pi1})-(\ref{eq:pi3}).
The myopic policy $g^m$ that is optimal for the finite horizon $T$ problem (by Theorem \ref{thm:myopic}) satisfies the dynamic program
\begin{align}
&J_{\beta,T}^{g^m}(\pi) = \nonumber\\
&\max_{n\in\{1,2,...,N\}}\left\lbrace \pi^n R 
                  + \beta E \{J_{\beta,T-1}^{g^m}(\pi_{1})|\pi_0=\pi,U_0=n\} \right\rbrace,
                  \label{infh:eq2}
\end{align}
where
\begin{align}
J_{\beta,T}^{g^m}(\pi) := & E^{g^m}\{\sum_{t=0}^{T} \beta^t R(t)|\pi_0=\pi\}.
\end{align}
Since the reward $R(t) \leq R_K$ is bounded, by the bounded convergence theorem we get
\begin{align}
J_\beta^{g^m}(\pi) = & E^{g}\{\sum_{t=0}^{\infty} \beta^t R(t)|\pi_0=\pi\} \nonumber\\
                   = & \lim_{T\rightarrow \infty}E^{g}\{
                             \sum_{t=0}^{T} \beta^t R(t)|\pi_0=\pi\} \nonumber\\          
                   = & \lim_{T\rightarrow \infty} J_{\beta,T}^{g^m}(\pi),
                   \label{infh:eq3}
\end{align} 
Letting $T\rightarrow\infty$ in (\ref{infh:eq2}) and using the bounded convergence theorem we obtain
\begin{align}
J_\beta^{g^m}(\pi) = &\max_{n \in \{1,2,...,N\}}\left\lbrace \pi^n R 
                  + \beta E \{J_{\beta}^{g^m}(\hat{\pi}(\pi,n))\} \right\rbrace,
                  \label{infh:eq4}
\end{align}
Notice that (\ref{infh:eq4}) is exactly the dynamic programming equation (\ref{infh:eq1}); therefore,
\begin{align}
J_\beta^{g^m}(\pi) = J_\beta^{g^*}(\pi);
\end{align}
consequently, the myopic policy $g^m$ is optimal for the infinite horizon problem (P2).
\end{myproof}

\section{Comparison with the Result of the Two-State Channel Model}
\label{sec:twostate}
The situation where each channel has two states, i.e. $K=2$, has been previously investigated in the literature (e.g. \cite{ahmad2009optimality}). In this section we show that when $K=2$ our conditions (A1)-(A4) reduce to the assumptions made in \cite{ahmad2009optimality}.
\\\\
When $K =2$, then $L$ has to be two, and the matrix of transition probabilities is given by
\begin{align}
&P_1 = (p_{1,1},p_{1,2}) = (1-p_{1,2},p_{1,2}),\\
&P_2 = (p_{2,1},p_{2,2}) = (1-p_{2,2},p_{2,2}).
\end{align}
In this case, for any two PMF $x,y \in \Pi$, let $x = (1-a,a),y=(1-b,b)$; then we have
\begin{align}
& x \geq_{st} y 
\Longleftrightarrow
 a \geq b.
\end{align}
Without loss of generality, let $R_1 = 0, R_2 = 1$, then our conditions 
reduce to the following conditions.
\\\\
For (A\ref{A:order}), we get
\begin{align}
&P_2 \geq_{st} P_1 \Longleftrightarrow
p_{2,2} \geq p_{1,2}
\label{A:orderfortwo}
\end{align}
For (A\ref{A:initial}) note that
\begin{align}
&\Pi = \{(1-p,p): 0\leq p \leq 1\};\\
&\Pi P = \{(1-p,p): p_{1,2}\leq p \leq p_{2,2}\}.
\end{align}
Consequently, (A\ref{A:initial}) reduces to
\begin{align} 
&\pi^n_0 = (1-p^n,p^n), p_{1,2}\leq p^n \leq p_{2,2}\text{ for } n=1,2,...,N (cf. (\ref{A:initial:eq1}) )\\
\text{and }& p^1 \leq p^2 \leq ... \leq p^N (cf. (\ref{A:initial:eq2})).
\end{align}
Using (\ref{A:orderfortwo}) we get
\begin{align}
P_1P =& p_{1,1}P_1+p_{1,2}P_2 \geq_{st} P_1,\\
P_2P =& p_{2,1}P_1+p_{2,2}P_2 \leq_{st} P_2,
\end{align}
thus (A\ref{A:K-1}) is automatically satisfied.
\\
For (A\ref{A:reward}), we have
\begin{align}
h = &  \frac{p_{2,2}-\beta p_{2,1}p_{1,2}}{1-\beta p_{2,1}}.
\end{align}
Therefore, 
\begin{align}
   \beta(h - P_1 R)
= & \beta\frac{p_{2,2}-p_{1,2}}{1-\beta p_{2,1}}
\leq  \frac{p_{2,2}-p_{1,2}}{ p_{2,2}}
\leq 	 1 = R_2-R_1.
\end{align}
Consequently, (A\ref{A:reward}) is automatically satisfied.
\\\\
As a result of the above analysis, our conditions (A1)-(A4) for the special case $K=2$ reduce to 
\begin{align}
& p_{2,2} \geq p_{1,2} \label{comp:eqp}\\
&\pi^n_0 = (1-p^n,p^n), p_{1,2}\leq p^n \leq p_{2,2}\text{ for } n=1,2,...,N\label{comp:eqinitial1}\\
& p^1 \leq p^2 \leq ... \leq p^N.\label{comp:eqinitial2}
\end{align}
Condition (\ref{comp:eqp}) is precisely the ``positively correlated" condition in \cite{ahmad2009optimality}.
Condition (\ref{comp:eqinitial1}) is satisfied, if the channels evolve before we begin sensing them (before time $t=0$).
Condition (\ref{comp:eqinitial2}) is always satisfied by renumbering of the channels.

\section{Myopic policy vs. Gittins index rule}
\label{sec:gittins}
In this section we investigate conditions under which the myopic policy coincides with the Gittins index rule.

Select a channel, say channel $n,n=1,2,...,N$. For PMF $\pi \in \Pi$, the Gittins index (\cite{gittins2011multi,gittins1979bandit}) of channel $n$ is defined is defined by
\begin{align}
\nu^n(\pi) := \max_{\tau} \frac{E^{g^{\tau}}[\sum_{t=0}^{\tau-1}\beta^t \pi^n_t R|\pi^n_0 = \pi]}
							{E^{g^{\tau}}[\sum_{t=0}^{\tau-1}\beta^t |\pi^n_0 = \pi]},
							\label{GI:eqGI}
\end{align}
where $\tau$ is any stopping time with respect to $\{\pi^n_t,t=0,1,...\} $ and $g^{\tau}$ chooses channel $n$ from $t=0$ up to $t=\tau-1$.
The Gittins index rule (\cite{gittins2011multi,gittins1979bandit}) chooses the channel with the highest Gittins index at every time instant $t$.

In condition (A\ref{A:K-1}) (Section \ref{sub:assum}) $L$ is fixed; it can be any number form $2$ to $K$. 
In this section we show that when $L=K$, under conditions (A1)-(A4),  after time $0$ the myopic policy coincides with the Gittins index rule. We establish this result via Theorem \ref{thm:GI} and \ref{thm:GIopt}.

\begin{thm}
\label{thm:GI}

\begin{enumerate}[(i)]
\item For $\pi \in \Pi P$, $P_{K-1} \leq_{st} \pi \leq_{st} P_{K}$, the Gittins index $\nu(\pi)$ is given by
\begin{align}
\nu(\pi) = \frac{\pi R + \beta \pi(K)\frac{P_K R}{1-\beta p_{KK}}}
							 {1 + \beta \pi(K)\frac{1}{1-\beta p_{KK}}}.
							 \label{GI:eqGIK}
\end{align}
\item If $ \pi_x,\pi_y \in \Pi P $, $P_{K-1} \leq_{st} \pi_y \leq_{st} \pi_x \leq_{st} P_{K}$, then 
$\nu(\pi_x)\geq \nu(\pi_y)$
\item If $\pi \in \Pi P$, $P_{K-1} \leq_{st} \pi \leq_{st} P_{K}$, then $\nu(\pi)\geq \nu(P_i)$ for $i<K$.
\end{enumerate}
\end{thm}
\begin{myproof}
(i). From Properties \ref{P:PMForder} and part (ii) of \ref{P:instreward} we know that
\begin{align}
\pi R \leq P_KR \text{ for all } \pi \in \Pi P.
\label{GI:equb1}
\end{align}
Using (\ref{GI:equb1}) in the definition of Gittins index (\ref{GI:eqGI}) we get
\begin{align}
\nu(\pi) \leq P_KR \text{ for all } \pi \in \Pi P.
\label{GI:equbGI}
\end{align}
Letting $\tau = 1$ in (\ref{GI:eqGI}), we get an lower bound on the Gittins index of $P_K$
\begin{align}
\nu(P_K) \geq E[R(\pi_0)|\pi_0 = P_K]=P_KR.
\label{GI:eqlbGI}
\end{align}
Combing (\ref{GI:equbGI}) and (\ref{GI:eqlbGI}) we obtain
\begin{align}
\nu(P_K) \geq P_KR \geq \nu(\pi) \text{ for all } \pi \in \Pi P.
\label{GI:eqQk}
\end{align}
Consequently, the PMF $P_K$ has the largest Gittins index among all PMFs.
\\
From Theorem 4.1 in \cite{varaiya1985extensions} we know that the second largest Gittens index 
among PMFs \\
$\{\pi, P_1,P_2,..,P_{K-1},P_K\}$ is given by
\begin{align}
\max_{x = \{\pi, P_1,P_2,..,P_{K-1}\}} \nu_K(x),
\end{align}
where
\begin{align}
\nu_K(x) :=& \frac{A_K(x)}{B_K(x)},\label{GI:nuK}\\
A_K(x) := &x R + \beta x(K)A_K(P_K), A_K(P_K)=\frac{P_KR}{1-\beta P_{KK}},\\
B_K(x) := &1 + \beta x(K)B_K(P_K), B_K(P_K)=\frac{1}{1-\beta P_{KK}}.
\end{align}
We now show that for $P_{K-1} \leq_{st} \pi \leq_{st} P_{K}$
\begin{align}
\nu_K(\pi)= \max_{x = \{\pi, P_1,P_2,..,P_{K-1}\}} \nu_K(x).
\end{align}
For that matter we need to show that $\nu(\pi_x) \geq \nu(\pi_y)$ whenever $\pi_x \geq_{st} \pi_y,\pi_x,\pi_y\in \Pi P$.
From (\ref{GI:nuK}),
\begin{align}
\nu_K(\pi_x) = & \frac{\pi_x R + \beta \pi_x(K)A_K(P_K)}{1 + \beta \pi_x(K)B_K(P_K)}\nonumber\\
         = & \frac{A_K(P_K)}{B_K(P_K)} + \frac{\pi_xR - \frac{A_K(P_K)}{B_K(P_K)}}{1 + \beta \pi_x(K)B_K(P_K)}\nonumber\\
         = & P_K R + \frac{\pi_xR - P_K R}{1 + \beta \pi_x(K)B_K(P_K)}\nonumber\\
 	  \geq & P_K R + \frac{\pi_yR - P_K R}{1 + \beta \pi_x(K)B_K(P_K)}\nonumber\\
 	  \geq & P_K R + \frac{\pi_yR - P_K R}{1 + \beta \pi_y(K)B_K(P_K)}\nonumber\\
 	     = & \nu_K(\pi_y) .\label{GI:eqcomp}
\end{align}
The first inequality in (\ref{GI:eqcomp}) follows from part (ii) of Property \ref{P:instreward} and  $\pi_x \geq_{st} \pi_y $.
The last inequality in (\ref{GI:eqcomp}) holds because $\pi_yR - P_K R \leq 0$ as $\pi_y\leq_{st} P_K $.\\
Since $\pi \geq_{st} P_{i}$ for $i=1,2,...,K-1$, (\ref{GI:eqcomp}) ensures that $\nu_K(\pi) \geq \nu_K(P_{i})$ for $i=1,2,...,K-1$.
Thus, $\pi$ is the PMF with the second largest Gittins index among $\{\pi, P_1,P_2,..,P_{K-1},P_K\}$.\\
The Gittins index for $\pi \in \Pi P, P_{K-1} \leq_{st} \pi \leq_{st} P_{K}$ is given by
\begin{align}
\nu(\pi) = \nu_K(\pi) = \frac{\pi R + \beta \pi(K)\frac{P_K R}{1-\beta p_{KK}}}
							 {1 + \beta \pi(K)\frac{1}{1-\beta p_{KK}}}.
							 \label{GI:eqGIforpi}
\end{align}
This completes the proof of (i).\\
(ii). If $ \pi_x,\pi_y \in \Pi P $, $P_{K-1} \leq_{st} \pi_y \leq_{st} \pi_x \leq_{st} P_{K}$,
by (\ref{GI:eqcomp}) and (\ref{GI:eqGIforpi}), we get
\begin{align}
\nu(\pi_y)=\nu_K(\pi_y)\leq \nu_K(\pi_x)= \nu(\pi_x ).
\end{align}
(iii). From part (i) we know that for $\pi \in \Pi P, P_{K-1} \leq_{st} \pi \leq_{st} P_{K}$, $\pi$ gives the second largest Gittins index among $\{\pi, P_1,P_2,..,P_{K-1},P_K\}$. Consequently, $\nu(\pi)\geq \nu(P_i)$ for $i<K$.

\end{myproof}
\begin{thm}
\label{thm:GIopt}
Under conditions (A1)-(A4) and $L=K$, after time $t=0$ the Gittins index rule is an optimal channel sensing policy for Problems (P1) and (P2).
\end{thm}
\begin{myproof}
Consider any time $t>0$.
If the channel observed at time $t-1$ is in state $K$ then the PMF of that channel at $t$ is $P_K$.
The myopic policy senses this channel at $t$. The Gittins index rule senses the same channel at $t$ as $P_K$ is the PMF with the largest Gittins index by Theorem \ref{thm:GI}, part (ii).\\
If the channel observed at time $t-1$ is in state $i,i<K$, then the PMF of that channel at $t$ is $P_i$
and the PMFs of all other channels are stochastically ordered and are stochastically larger than $P_{K-1}$ and 
stochastically smaller than $P_K$ by Property \ref{P:PMForder}.
The myopic policy will choose the channel with the stochastically largest PMF (among all channels that are not observed at $t-1$). By Theorem \ref{thm:GI} (ii), the Gittins index of the same channel is the largest among the Gittins indices of all channels that are not observed at $t-1$. 
By Theorem \ref{thm:GI} (iii), the Gittins index of the channel observed at time $t-1$ is $\nu(P_i) \leq \nu(\pi)$ for all $P_{K-1} \leq_{st} \pi \leq_{st} P_{K}$. Therefore, the Gittins index chooses the same channel as the myopic policy.
From the optimality of the myopic policy, under conditions (A1)-(A4) (Theorem \ref{thm:myopic} and \ref{thm:myopic2}) and the condition $L=K$, after time $t=0$ the Gittins index rule is an optimal channel sensing strategy for problem (P1) and (P2).
\end{myproof}
Note that, if two channels, say channel $1$ and $2$ are such that $\pi^1_0,\pi^2_0 \in \{P_1,P_2,...,P_{K-1}\}$ then
$\pi^1_0,\pi^2_0 \in \Pi P$ and thus, (A\ref{A:initial}) is satisfied. Nevertheless 
$\pi^1_0,\pi^2_0$ do not necessarily satisfy the condition $P_{k-1}\leq_{st}\pi^i_0\leq_{st}P_K $ of Theorem \ref{thm:GI}.
Thus, at $t=0$, the assertion of Theorem \ref{thm:GI} may not be true for channels $1$ and $2$, thus the Gittins index rule may not be optimal  at time $0$.

\section{Conclusion}
\label{sec:conclusion}
We investigated a channel sensing problem where each channel has more than two states. 
We formulated an optimization problem which is an instance of the restless bandit problem. 
For this problem, we identified conditions sufficient to guarantee the optimality of the myopic policy, the policy that selects at each time instant the channel with the stochastically largest PMF on its states. 
We also identified conditions under which the Gittins index rule coincides with the myopic policy (and is optimal).

Our results on the optimality of the myopic policy extend previously existing results on the same problem when each channel has two states. In our opinion such an extension is non-trivial for the following reason.
When each channel has two states, the information states of the channels can always be totally ordered (as each information state is described by a single number); on the other hand, when each channel has more than two states, the information states of the channels (expressed by their PMF on the states) are not even guaranteed to be partially ordered. Such a lack of order creates serious technical problems, and requires significant insight into the nature of the problem (so as to identify the appropriate assumptions), and much more careful and complicated analysis (so as to establish the optimality of the myopic policy).

Our results on the optimality of the Gittins index rule rely on : 
(i) the fact that the information state of any channel after $t>0$ lies stochastically between $P_{K-1}$ and $P_K$, i.e. $P_{K-1}\leq_{st} \pi \leq_{st} P_K$; and (ii) the fact that 
$\nu(\hat{\pi})\geq \nu(\pi)$ whenever $\hat{\pi}\geq_{st} \pi$ and both $\hat{\pi}$ and $\pi$ are stochastically ordered between $P_{K-1}$ and $P_K$. 
We have not been able to prove whether or not the Gittins index rule coincides with the myopic policy when conditions (A1)-(A4) are valid and $L\neq K$ in (A\ref{A:K-1}).


%

\appendices
\section{}
\label{app:pmf}
\begin{myproof}[\text{Proof of Property \ref{P:preserveorder}} ]

 \label{pfP:preserveorder} 

\begin{align}
xP-yP = &\sum_{i=1}^K (x(i)-y(i))P_i \nonumber\\
      = &\sum_{i=2}^K \left[\left(\sum_{j=i}^K(x(j)-y(j))\right)(P_i-P_{i-1})\right]. \label{pfP1:eq1}
\end{align}
The last equality follows from a standard identity on the summation by parts of two sequence
$\{(x(i)-y(i)),i=1,2,...,K\}$ and $\}P_i,i=1,2,...,K\}$.
Note that $\sum_{j = i}^K(x(j)-y(j)) \geq 0$ since $x \geq_{st} y$, and by assumption (A\ref{A:order}) $P_i \geq_{st} P_{i-1}$.\\
Consequently, $\left(\sum_{j=i}^K(x(j)-y(j))\right)(P_i-P_{i-1}) \geq_{st} \mathbf{0}$, 
where $\mathbf{0}:= (0,0,...,0)$ is the zero vector.
Thus by (\ref{pfP1:eq1}) 
\begin{align}
xP-yP \geq_{st} &\sum_{i=1}^K \mathbf{0} = \mathbf{0},
\end{align}
Hence, $xP \geq_{st} yP$.
\end{myproof}

\begin{myproof}[Proof of Property \ref{P:PMForder} ] \label{pfP:PMForder}
\begin{align}
xP^2 = \sum_{i=1}^K  x(i)P_iP
\label{PfP:PMForder:eq1}
\end{align}
Then, from Property \ref{P:preserveorder} , (A\ref{A:order}) and (A\ref{A:K-1}) we obtain
\begin{align}
& P_iP  \leq_{st} P_KP \leq_{st} P_L \label{PfP:PMForder:eq2}\\
& P_iP  \geq_{st} P_1P \geq_{st} P_{L-1}\label{PfP:PMForder:eq3}
\end{align}
The first inequality in (\ref{PfP:PMForder:eq2}) and the first inequality in (\ref{PfP:PMForder:eq3})
are true because of Property \ref{P:preserveorder} and the fact that $P_1\leq_{st} P_i \leq_{st} P_K$ (condition (A\ref{A:order})).
The second inequality in (\ref{PfP:PMForder:eq2}) and the second inequality in (\ref{PfP:PMForder:eq3})
are true because of condition (A\ref{A:K-1}).\\
Therefore, (\ref{PfP:PMForder:eq1}) along with (\ref{PfP:PMForder:eq2}) and (\ref{PfP:PMForder:eq3}) give
\begin{align}
P_{L-1} \leq_{st} xP^2 \leq_{st} P_L
\end{align}
\end{myproof}

\begin{myproof}[Proof of Property \ref{P:canbeorder} ] \label{pfP:canbeorder}
We prove this Property by induction. The Property is true at $t=0$ by (A\ref{A:initial}). 
\\
Now assume the Property is true at $t$.\\
If $n,m$ are not selected at $t$, $\pi^n_{t+1} = \pi^n_{t}P$, $\pi^m_{t+1} = \pi^m_{t}P$.\\
By the induction hypothesis we have $\pi^n_t \leq_{st} \pi^m_t$ or $\pi^m_t \leq_{st} \pi^n_t$. 
Then by Property \ref{P:preserveorder}, we obtain 
$\pi^n_tP \leq_{st} \pi^m_tP$ or $\pi^m_tP \leq_{st} \pi^n_tP$
, consequently, $\pi^n_{t+1} \leq_{st} \pi^m_{t+1}$ or $\pi^m_{t+1} \leq_{st} \pi^n_{t+1}$. 
\\
Suppose, without loss of generality, that channel $n$ is selected at $t$.\\
Since channel $m$ is not selected at $t$, $\pi^m_{t+1} = \pi^m_{t}P \in \Pi P^2$.\\
If the observed state is $i \geq L$, then by Property \ref{P:PMForder}, $\pi^n_{t+1}=P_i \geq_{st} P_L \geq_{st} \pi^m_{t+1}$.\\
If the observed state is $i <L$, then, again by Property \ref{P:PMForder}, $\pi^n_{t+1}=P_i \leq_{st}P_{L-1} \leq_{st} \pi^m_{t+1}$. Consequently, $\pi^n_{t+1} \leq_{st} \pi^m_{t+1}$ or $\pi^m_{t+1} \leq_{st} \pi^n_{t+1}$. 
\end{myproof}

\section{}
\label{app:inst}
\begin{myproof}[Proof of Property \ref{P:instreward}]
\label{pfP:instreward}
\begin{enumerate}[(i)]
\item By summation by parts we have
\begin{align}
(x-y)v = &\sum_{i=1}^K (x(i)-y(i))v_i\nonumber\\
       = &\sum_{i=2}^K \left[\left(\sum_{j=i}^K(x(j)-y(j))\right)(v_i-v_{i-1})\right]. \label{pflm:vec:eq1}
\end{align}
Since $x \geq_{st} y$, 
\begin{align}
\sum_{j = i}^K(x(j)-y(j)) \geq 0.\label{pflm:vec:eq2}
\end{align}
The condition $v_i \geq v_{i-1}$, $i=2,3,...,K-1$ in the statement of Property \ref{P:instreward}, and
(\ref{pflm:vec:eq2}) give
\begin{align}
\left(\sum_{j=i}^K(x(j)-y(j))\right)(v_i-v_{i-1}) \geq 0 
\text{ for all }i = 2,3,...,K.
\label{pflm:vec:eq3}
\end{align}
Then (\ref{pflm:vec:eq3}) and (\ref{pflm:vec:eq1}) result in
\begin{align}
(x-y)v \geq 0.
\label{lmvec}
\end{align}
\item From the definition of $U$ we have:
\begin{align}
&\text{For }i <L, U_i - U_{i-1} = R_i - R_{i-1}.\label{pfP:instreward:eq01}\\
&\text{For }i \geq L, U_i - U_{i-1} = R_i - R_{i-1} + \beta(P_i-P_{i-1})U \geq R_i - R_{i-1}.\label{pfP:instreward:eq02}
\end{align}
Then, for all $i$, by the definition of $M$ we obtain
\begin{align}
M_i - M_{i-1} = &U_i - U_{i-1} +\sum_{i \geq L}p_{Ki}(P_i-P_{i-1})U \nonumber\\
             \geq &U_i - U_{i-1} \nonumber\\
             \geq &R_i - R_{i-1} \geq 0.
 \label{pfP:instreward:eq1}
\end{align}
The first inequality in (\ref{pfP:instreward:eq1}) holds because of condition (A4) (eq. (\ref{A:reward:eq1})). The second inequality 
in (\ref{pfP:instreward:eq1}) follows from (\ref{pfP:instreward:eq01}) and (\ref{pfP:instreward:eq02}).
From (\ref{pfP:instreward:eq1}), it follows that
 $M-U$ and $U-R$ are in increasing order (i.e. $M_i-U_i$ and $U_i-R_i$ increase as $i$ increases).\\
Since $x\geq_{st}y$, from (\ref{pfP:instreward:eq1}) and the result of part (i) we have
\begin{align}
(x-y)M \geq (x-y)U \geq (x-y)R \geq 0.
\end{align}
\item 
Because of Assumption (A\ref{A:reward}) and the result of part (ii) we have: 
\begin{align}
&\text{For }i <L, U_i - U_{i-1} = R_i - R_{i-1} \geq \beta(P_i-P_{i-1})M \geq \beta(P_i-P_{i-1})U.\label{pfP:instreward:eq2}\\
&\text{For }i \geq L, U_i - U_{i-1} = R_i - R_{i-1} + \beta(P_i-P_{i-1})U \geq \beta(P_i-P_{i-1})U.\label{pfP:instreward:eq3}
\end{align}
Then, (\ref{pfP:instreward:eq2}) and (\ref{pfP:instreward:eq3}) imply that $U-\beta PU$ is in increasing order, consequently by the result of part (i) we obtain
\begin{align}
(x-y)U \geq \beta (x-y)PU. \label{pfP:instreward:eq4}
\end{align}
Since $M = U + \beta\sum_{i\geq L} p_{Ki} PU$, 
\begin{align}
(x-y)M =& (x-y)(U + \beta\sum_{i\geq L} p_{Ki} PU) \nonumber\\
       =& (x-y)U + \beta\sum_{i\geq L} p_{Ki} (xP-yP)U \nonumber\\
   \geq & \beta(x-y)PU + \beta\sum_{i\geq L} p_{Ki} \beta(xP-yP)PU \nonumber\\
       =& \beta (x-y)PM, \label{pfP:instreward:eq5}
\end{align}
where the inequality in (\ref{pfP:instreward:eq5}) is a consequence of (\ref{pfP:instreward:eq4}).
\item 
If $x(i)=y(i)$ for all $i \geq L$, then $x(i)-y(i)=0$ for $i \geq L$.\\
Define $v :=(v_1,v_2,...,v_K)$ such that
\begin{align}
&v_i = R_i-\beta P_iM \text{ for } i=1,2,...,L-1,\label{pfP:instreward:eqvi1}\\
&v_i = v_{L-1} \text{ for } i\geq L.\label{pfP:instreward:eqvi2}
\end{align}
From assumption (\ref{A:reward:eq1}) in (A\ref{A:reward}) we know that 
$v_i-v_{i-1} = R_i-R_{i-1}-\beta (P_i-P_{i-1})M \geq 0$ for $i \leq L-1$ and $v_i-v_{i-1} =0$ for $i\geq L$.
Then by the result of part (i) we obtain
\begin{align}
(x-y)(R - \beta PM) =& \sum_{i=1}^{L-1} (x(i)-y(i))(R_i-\beta P_iM) \nonumber\\
					=& \sum_{i=1}^{L-1} (x(i)-y(i))v_i + \sum_{i\geq L} (x(i)-y(i))v_i\nonumber\\
					=& (x-y)v \geq 0.\label{pfP:instreward:eq8}
\end{align}
The second equality in (\ref{pfP:instreward:eq8}) follows from the definition of $v_i$ (eq. (\ref{pfP:instreward:eqvi1})) and the fact that $x(i)-y(i)=0$ for $i\geq L$.
The inequality in (\ref{pfP:instreward:eq8}) is true by the result of part (i).
\\
Since $M = U + \beta\sum_{i\geq L} p_{Ki} PU$ and $x\geq_{st} y$, it follows that
\begin{align}
\beta(x-y)PU \leq  \beta(x-y)P(U+\beta\sum_{i\geq L} p_{Ki}PU) =\beta(x-y)PM  \leq (x-y)R, \label{pfP:instreward:eq9}
\end{align}
where the first inequality in (\ref{pfP:instreward:eq9}) follows from the fact that $xP^2\geq_{st}yP^2$, the fact that $U_i$ is increasing with $i$, and the result of part (i); and the last inequality in (\ref{pfP:instreward:eq9}) follows from (\ref{pfP:instreward:eq8}).\\
The case where $x(i)=y(i)$ for all $i < L$ can be proved in the same way.
\end{enumerate}
\end{myproof}
\section{}
\label{app:fun}
\begin{myproof}[Proof of Property \ref{P:myopic}]
We want to show that under $g^{O_0}_{0:T}$, at any time $t$ the ordering $O_t$ 
has the property that\\
$\pi^{O_t(1)}_t\leq_{st}\pi^{O_t(2)}_t\leq_{st}...\leq_{st}\pi^{O_t(N)}_t$.\\
At $t=0$, by the statement of Property \ref{P:myopic}, the initial ordering $O_0$ is such that $\pi^{O_0(1)}_0\leq_{st}\pi^{O_0(2)}_0\leq_{st}...\leq_{st}\pi^{O_0(N)}_0$.\\
Suppose at time $t$, the ordering $O_t$ is such that
$\pi^{O_t(1)}_t\leq_{st}\pi^{O_t(2)}_t\leq_{st}...\leq_{st}\pi^{O_t(N)}_t$.\\
If the observation is $Y_t \geq L$, the new ordering is
$O_{t+1}=\hat{m}(O_t,Y_t)=O_t$
and the PMFs of the channels evolves to 
\begin{align}
&\pi^{n}_{t+1} = \pi^{n}_tP \text{ for } n\neq O_t(N),\\
&\pi^{O_t(N)}_{t+1} = P_{Y_t} \geq_{st} P_L.
\end{align}
From Properties \ref{P:preserveorder} and \ref{P:PMForder} we know that 
\begin{align}
&\pi^{O_t(1)}_tP\leq_{st}\pi^{O_t(2)}_tP\leq_{st}...\leq_{st}\pi^{O_t(N-1)}_tP\leq_{st} P_{L} \leq_{st} P_{Y_t} ,
\end{align}
therefore,
\begin{align}
&\pi^{O_{t+1}(1)}_{t+1}\leq_{st}\pi^{O_{t+1}(2)}_{t+1}\leq_{st}...\leq_{st}\pi^{O_{t+1}(N)}_{t+1}.
\end{align}
On the other hand, if the observation is $Y_t < L$, the new ordering is
$O_{t+1}=\hat{m}(O_t,Y_t)=SO_t$
and the PMFs of the channels become
\begin{align}
&\pi^{n}_{t+1} = \pi^{n}_tP \text{ for } n\neq O_t(N),\\
&\pi^{O_t(N)}_{t+1} = P_{Y_t} \leq_{st} P_{L-1}.
\end{align}
Again, from Properties \ref{P:preserveorder} and \ref{P:PMForder} we get 
\begin{align}
&P_{Y_t} \leq_{st} P_{L-1} \leq_{st} \pi^{O_t(1)}_tP\leq_{st}\pi^{O_t(2)}_tP\leq_{st}...\leq_{st}\pi^{O_t(N-1)}_tP, 
\end{align}
hence,
\begin{align}
&\pi^{O_{t+1}(1)}_{t+1}\leq_{st}\pi^{O_{t+1}(2)}_{t+1}\leq_{st}...\leq_{st}\pi^{O_{t+1}(N)}_{t+1}.
\end{align}
Thus, the ordering-based policy $g^{O_0}_{0:T}$
selects at any time $t$ the channel $O_t(N)$ from the ordering $O_t$
with $\pi^{O_t(1)}_t\leq_{st}\pi^{O_t(2)}_t\leq_{st}...\leq_{st}\pi^{O_t(N)}_t$.
This ordering-based policy is exactly the same as the myopic policy $g^m$.
\end{myproof}

\section{}
We first establish a lemma that is needed for the proof of Properties 
\ref{P:funrotate}-\ref{P:funlift}.

\begin{lemma} \label{lm:funlin} 
The functions $V_t(O_t,\pi^1_t,\pi^2_t,...,\pi^N_t)$, $t=1,2,...,T$ (defined by eq. (\ref{eqVtdef})), are linear in every component $\pi^n_t,n=1,2,...,N$.\\
That is, for all $n = 1,2,...,N$
\begin{align}
V_t(O_t,\pi^1_t,\pi^2_t,...,\pi^N_t)
=& \sum_{i = 1}^K \pi^n(i) V_t(O_t,\pi^1_t,...,\pi^{n-1}_t,e_i,\pi^{n+1}_t,...,\pi^N_t),
\end{align}
where $e_i$ is the vector with $1$ in the $ith$ position and $0$ otherwise, i.e.
$\begin{array}{rl}
e_i = [0,...,0,&1,0,...,0] \\ 
 &\uparrow i\text{th position}
\end{array}$.\\
Furthermore, $L_t(O_t,\hat{\pi}^1_t,\pi^1_t,\pi^2_t,...,\pi^N_t)$ satisfies for $n = 2,3,...,N$
\begin{align}
L_t(O_t,\hat{\pi}^1_t,\pi^1_t,\pi^2_t,...,\pi^N_t)
=& \sum_{i = 1}^K \pi^n(i) L_t(O_t,\hat{\pi}^1_t,\pi^1_t,...,\pi^{n-1}_t,e_i,\pi^{n+1}_t,...,\pi^N_t), 
\label{lm:funlin:eqL1}
\\
L_t(O_t,\hat{\pi}^1_t,\pi^1_t,\pi^2_t,...,\pi^N_t)
=& \sum_{i = 1}^K (\hat{\pi}^1_t(i)-\pi^1_t(i)) 
V_t(O_t,e_i,\pi^2_t,...,\pi^N_t).
\label{lm:funlin:eqL2}
\end{align}

\end{lemma}	
\begin{myproof} \label{pflm:expandei}By definition of $V_t$ (eq (\ref{eqVtdef})) we have
\begin{align}
V_t(O_t,\pi^1_t,\pi^2_t,...,\pi^N_t) 
:=& E^{g^{O_t}_{t:T}}[\sum_{s=t}^T \beta^{s-t}R(s)
| \pi^1_t,\pi^2_t,...,\pi^N_t  ]  \nonumber\\
= &\sum_{i=1}^K \pi^n_t(i) 
E^{g^{O_t}_{t:T}}[\sum_{s=t}^T \beta^{s-t}R(s)| \pi^1_t,\pi^2_t,...,\pi^N_t, X^{n}_t=i ].
\label{pflm:expandei:eq1}
\end{align}
Because of the specification of the ordering-based policy $g^{O_t}_{t:T}$ and the fact that conditional on $\{X^{n}_t=i,\pi^{n}_t\}$ the evolution of channel $n$ is the same as that conditional on
$\{\pi^{n}_t=e_i\}$, we have
\begin{align}
&E^{g^{O_t}_{t:T}}[\sum_{s=t}^T \beta^{s-t}R(s)| \pi^1_t,\pi^2_t,...,\pi^N_t, X^{n}_t=i ]\nonumber\\
= &E^{g^{O_t}_{t:T}}[\sum_{s=t}^T \beta^{s-t}R(s)| \pi^{1}_t,...,\pi^{n-1}_t, \pi^{n+1}_t,...,\pi^{N}_t,\pi^{n}_t=e_i].\label{pflm:expandei:eq2}
\end{align}
Then from (\ref{pflm:expandei:eq1}) and (\ref{pflm:expandei:eq2}) we obtain
\begin{align}
&V_t(O_t,\pi^1_t,\pi^2_t,...,\pi^N_t) \nonumber\\
= &\sum_{i=1}^K \pi^n_t(i) 
E^{g^{O_t}_{t:T}}[\sum_{s=t}^T \beta^{s-t}R(s)| \pi^{1}_t,...,\pi^{n-1}_t, \pi^{n+1}_t,...,\pi^{N}_t,\pi^{n}_t=e_i]\nonumber\\
= &\sum_{i = 1}^K \pi^n_t(i) V_t(O_t,\pi^1_t,...,\pi^{n-1}_t,e_i,\pi^{n+1}_t,...,\pi^N_t).
\end{align}
Furthermore, $L_t$ is the difference of two $V_t$'s, so the linearity of $V_t$ leads directly to equations (\ref{lm:funlin:eqL1}) and (\ref{lm:funlin:eqL2}).
\end{myproof}
\noindent
We Proceed now with the proof of Properties \ref{P:funrotate}-\ref{P:funlift}.
In the following proofs, we use the notation
\begin{align}
&\pi^{k_1:k_2}_t:=(\pi^{k_1}_t,\pi^{k_1+1}_t,...,\pi^{k_2}_t)\\
&\pi ^{k_1:k_2}_tP:=(\pi^{k_1}_tP,\pi^{k_1+1}_tP,...,\pi^{k_2}_tP)
\end{align}

\begin{myproof} [Proof of Properties \ref{P:funrotate}-\ref{P:funlift}]\label{pfP:fun}
First note that Property \ref{P:funswitch2} is a special case of Property \ref{P:funswitch}. This can be seen as follows.\\
Without loss of generality, let $O_t(n)=1,O_t(m)=2$, and $\pi^{1}_t \geq_{st}\pi^2_t $.
Note that
\begin{align}
V_t(O_t,\pi^2_t,\pi^2_t,...,\pi^N_t) =V_t(W_{nm}O_t,\pi^2_t,\pi^2_t,...,\pi^N_t).
\label{pfP:funswitch2:eq1}
\end{align}
Applying Property \ref{P:funswitch} at time $t$, we have
\begin{align}
&V_t(O_t,\pi^1_t,\pi^2_t,...,\pi^N_t) -
V_t(W_{nm}O_t,\pi^1_t,\pi^2_t,...,\pi^N_t) \nonumber\\
=& 
 V_t(O_t,\pi^1_t,\pi^2_t,...,\pi^N_t) - V_t(O_t,\pi^2_t,\pi^2_t,...,\pi^N_t)  
+V_t(W_{nm}O_t,\pi^2_t,\pi^2_t,...,\pi^N_t) -V_t(W_{nm}O_t,\pi^1_t,\pi^2_t,...,\pi^N_t) \nonumber\\
= & L_t(O_t,\pi^1_t,\pi^2_t,\pi^2_t,...,\pi^N_t)
- L_t(W_{nm}O_t,\pi^1_t,\pi^2_t,\pi^2_t,...,\pi^N_t)
\geq 0.
\label{pfP:funswitch2:eq2}
\end{align}
The first equality in (\ref{pfP:funswitch2:eq2}) holds because of (\ref{pfP:funswitch2:eq1}).
The second equality is a consequence of the definition of $L_t$ (eq (\ref{eq:L})).
The inequality follows from Property \ref{P:funswitch} at $t$.
\\
Therefore, Property \ref{P:funswitch2} is true at time $t$ once Property \ref{P:funswitch} is true at time $t$.

We will prove all three Properties \ref{P:funrotate}, \ref{P:funswitch} and \ref{P:funlift} simultaneously by induction.
\\
We remind the reader that for Properties \ref{P:funrotate}, \ref{P:funswitch} and \ref{P:funlift}
$O_t\in \mathcal{O}$ with $O_t(n)=1$, $1\leq m<n\leq N $ and 
\begin{align}
&S^{-m}O_t =  (O_t(m+1),O_t(m+2),...,O_t(N),O_t(1),...,O_t(m)),\\
&W_{nm}O_t(i) =\left\lbrace
\begin{array}{ll}
O_t(n) & \text{ for }i=m\\
O_t(m) & \text{ for }i=n\\
O_t(i) & \text{otherwise}
\end{array}
\right. ,\\
&A_{nm} O_t(i) =\left\lbrace
\begin{array}{ll}
O_t(n) & \text{ for }i=m\\
O_t(i-1) & \text{ for }i=m+1,m+2,...,n\\
O_t(i) & \text{otherwise}
\end{array}
\right. .
\end{align}
For both the basis of induction and the induction we consider two cases. 
\begin{enumerate}[(i)]
\item When channel $1$ is not the right-most channel in $O_t$ (i.e. $n\neq N$ and $O_t(N)\neq 1$).
\item When channel $1$ is the right-most channel in $O_t$ (i.e. $n= N$ and $O_t(N)= 1$).
\end{enumerate}
\underline{\textbf{Basis of induction}}\\
For Property \ref{P:funrotate}:\\ 
(i) If $O_T(N)\neq 1$ (i.e. $n\neq N$),
\begin{align}
L_T(O_T,\hat{\pi}^1_T,\pi^{1:N}_T) -
L_T(S^{-m}O_T,\hat{\pi}^1_T,\pi^{1:N}_T) = 
(\pi^{O_T(N)}_TR-\pi^{O_T(N)}_TR) - (\pi^{O_T(m)}_TR-\pi^{O_T(m)}_TR) =0.
\end{align}
(ii) If $O_T(N)= 1$ (i.e. $n = N$), then
\begin{align}
L_T(O_T,\hat{\pi}^1_T,\pi^{1:N}_T) -
L_T(S^{-m}O_T,\hat{\pi}^1_T,\pi^{1:N}_T) = 
(\hat{\pi}^{1}_TR-\pi^{1}_TR) - (\pi^{O_T(m)}_TR-\pi^{O_T(m)}_TR) 
=(\hat{\pi}^{1}_T-\pi^{1}_T)R.
\label{pfP:basis:eq1}
\end{align}
By part (ii) of Property \ref{P:instreward} and $\hat{\pi}^1_t \geq_{st}\pi^1_t$ we get
\begin{align}
(\hat{\pi}^1_T-\pi^1_T) U \geq  (\hat{\pi}^1_T-\pi^1_T)R \geq  0.	
\label{pfP:basis:eq2}
\end{align}
Combing (\ref{pfP:basis:eq1}) with (\ref{pfP:basis:eq2}) we obtain
\begin{align}
(\hat{\pi}^1_T-\pi^1_T) U \geq  (\hat{\pi}^1_T-\pi^1_T)R =L_T(O_T,\hat{\pi}^1_T,\pi^{1:N}_T) -
L_T(S^{-m}O_T,\hat{\pi}^1_T,\pi^{1:N}_T) \geq  0.	
\end{align}
For Property \ref{P:funswitch}:\\ 
(i) If $O_T(N)\neq 1$ (i.e. $n\neq N$),
\begin{align}
L_T(O_T,\hat{\pi}^1_T,\pi^{1:N}_T) -
L_T(W_{nm}O_T,\hat{\pi}^1_T,\pi^{1:N}_T) = 
(\pi^{O_T(N)}_TR-\pi^{O_T(N)}_TR) - (\pi^{O_T(m)}_TR-\pi^{O_T(m)}_TR) =0.
\end{align}
(ii) If $O_T(N)= 1$ (i.e. $n = N$), then
\begin{align}
L_T(O_T,\hat{\pi}^1_T,\pi^{1:N}_T) -
L_T(W_{nm}O_t,\hat{\pi}^1_t,\pi^{1:N}_T) = 
(\hat{\pi}^{1}_TR-\pi^{1}_TR) - (\pi^{O_T(m)}_TR-\pi^{O_T(m)}_TR) 
=(\hat{\pi}^{1}_T-\pi^{1}_T)R.
\label{pfP:basis:eq3}
\end{align}
By part (ii) of Property \ref{P:instreward} and $\hat{\pi}^1_t \geq_{st}\pi^1_t$ we get
\begin{align}
(\hat{\pi}^1_T-\pi^1_T) M \geq  (\hat{\pi}^1_T-\pi^1_T)R \geq  0.	
\label{pfP:basis:eq4}
\end{align}
Combing (\ref{pfP:basis:eq3}) with (\ref{pfP:basis:eq4}) we obtain
\begin{align}
(\hat{\pi}^1_T-\pi^1_T) M \geq  (\hat{\pi}^1_T-\pi^1_T)R =L_T(O_T,\hat{\pi}^1_T,\pi^{1:N}_T) -
L_T(S^{-m}O_t,\hat{\pi}^1_t,\pi^{1:N}_T) \geq  0.	
\end{align}
For Property \ref{P:funlift}:\\
Since $P_K\geq P_i$, by part (ii) of Property \ref{P:instreward}, we get
\begin{align}
h := \frac{P_K R - \beta\sum_{i<L}p_{Ki}P_iR}{1-\beta\sum_{i<L}p_{Ki}}
\geq \frac{P_K R - \beta\sum_{i<L}p_{Ki}P_KR}{1-\beta\sum_{i<L}p_{Ki}}
= P_KR
\end{align}
Consequently, part (ii) of Property \ref{P:instreward} ensures that 
\begin{align}
\pi R \leq P_K R \leq h \text{ for all }\pi\in \Pi P.
\label{pfP:basis:eqh}
\end{align}
Then:\\
(i) If $O_T(N)\neq 1$ (i.e. $n\neq N$), we have
\begin{align}
& V_T(A_{nm}O_T,\pi^{1:N}_T) -
V_t(O_T,\pi^{1:N}_T)
= \pi^{O_T(N)}_T R-\pi^{O_T(N)}_TR = 0 \leq  h-\pi^{1}_TP^{N-n} R.
\label{pfP:basis:eq91}
\end{align}
The inequality in (\ref{pfP:basis:eq91}) follows from (\ref{pfP:basis:eqh}) and the fact that $\pi^{1}_TP^{N-n}\in \pi P$.
\\
(ii) If $O_T(N)= 1$(i.e. $n = N$), we have
\begin{align}
V_T(A_{nm}O_T,\pi^{1:N}_T) -
V_t(O_T,\pi^{1:N}_T) 
= & \pi^{O_T(N-1)}_T R-\pi^{1}_TR \leq  h-\pi^{1}_T R.
\label{pfP:basis:eq92}
\end{align}
The inequality in (\ref{pfP:basis:eq92}) follows from (\ref{pfP:basis:eqh}).
\\
This completes the basis of induction.
\\\\
\underline{\textbf{Induction hypothesis}}\\
Assume that the assertions of Properties \ref{P:funrotate}, \ref{P:funswitch} and \ref{P:funlift} are true for 
time $t+1,t+2,...,T$. \\
\underline{\textbf{Induction step}}\\
We prove here Properties \ref{P:funrotate}, \ref{P:funswitch} and \ref{P:funlift} for $t$.
\\
We first develop five expressions 
(\ref{pfP:funrotate:eq1notend}),(\ref{pfP:funrotate:eqorder1}),(\ref{pfP:funrotate:eqorder2}),
(\ref{pfP:funrotate:eq1isend}) and (\ref{pfP:funrotate:eqU})
 for $L_t$ and $L_{t+1}$ defined by eq. (\ref{eq:L}), that will be useful in the sequel.
\\
For any PMF $\pi\in \Pi $ we define
\begin{align}
&\underline{\pi} := (\pi(1),\pi(2),...,\pi(L-2),\sum_{i=L-1}^K \pi(i),0,...,0),
\label{pfP:undxdef}
\\
&\bar{\pi} := (0,...,0,\sum_{i=1}^L \pi(i),\pi(L+1),...,\pi(K))
\label{pfP:barxdef}
\end{align}
Then, $\underline{\pi}, \bar{\pi} \in \Pi$, and
\begin{align}
\pi = \underline{\pi}+\bar{\pi} - e_L + \sum_{i=L}^K\pi(i)(e_L-e_{L-1})
\label{pfP:underbarx}
\end{align}
Furthermore, if $\hat{\pi} \geq_{st} \pi$, it follows that
\begin{align}
&\underline{\hat{\pi}} \geq_{st} \underline{\pi},\\
&\bar{\hat{\pi}} \geq_{st} \bar{\pi}.
\end{align}
Consider any arbitrary ordering $O \in \mathcal{O}$. 
When $O(N) \neq 1$, assume $O(N)=2$ without any loss of generality. Then,
\begin{align}
&L_t(O,\hat{\pi}^1_t,\pi^1_t,\pi^{2:N}_t)\nonumber\\
:= &V_t(O,\hat{\pi}^1_t,\pi^{2:N}_t)
-V_t(O,\pi^1_t,\pi^{2:N}_t)\nonumber\\
= &(\pi^{2}_tR-\pi^{2}_tR) 
+\beta\sum_{i<L} \pi^2_t(i)
( V_{t+1}(SO,\hat{\pi}^1_tP,P_i,\pi^{3:N}_tP)-V_{t+1}(SO,\pi^1_tP,P_i,\pi^{3:N}_tP) )\nonumber\\
&\qquad +\beta\sum_{i\geq L} \pi^2_t(i)
( V_{t+1}(O,\hat{\pi}^1_tP,P_i,\pi^{3:N}_tP)-V_{t+1}(O,\pi^1_tP,P_i,\pi^{3:N}_tP) )\nonumber\\
= &\beta\sum_{i<L} \pi^2_t(i)
 L_{t+1}(SO,\hat{\pi}^1_tP,\pi^1_tP,P_i,\pi^{3:N}_tP)+\beta\sum_{i\geq L} \pi^2_t(i)
L_{t+1}(O,\hat{\pi}^1_tP,\pi^1_tP,P_i,\pi^{3:N}_tP).
\label{pfP:funrotate:eq1notend}
\end{align}
The second equality in (\ref{pfP:funrotate:eq1notend}) follows from the recursive equation for $V_t$ (eq. (\ref{eqVt})).
The last equality in (\ref{pfP:funrotate:eq1notend}) follows from the definition of $L_t$ (eq. \ref{eq:L}).
\\
Furthermore, by the induction hypothesis for Property \ref{P:funrotate}, we get, for all $i=1,2,...,K$,
\begin{align}
L_{t+1}(SO,\hat{\pi}^1_tP,\pi^1_tP,P_i,\pi^{3:N}_tP)
\geq L_{t+1}(O,\hat{\pi}^1_tP,\pi^1_tP,P_i,\pi^{3:N}_tP).
\label{pfP:funrotate:eq2}
\end{align}
Therefore,
\begin{align}
 &\beta L_{t+1}(SO,\hat{\pi}^1_tP,\pi^{1:N}_tP)\nonumber\\
=& \beta\sum_{i=1}^{L} \pi^2_t(i)L_{t+1}(SO,\hat{\pi}^1_tP,\pi^1_tP,P_i,\pi^{3:N}_tP)\nonumber\\
\geq &\beta\sum_{i<L} \pi^2_t(i)
 L_{t+1}(SO,\hat{\pi}^1_tP,\pi^1_tP,P_i,\pi^{3:N}_tP)+\beta\sum_{i\geq L} \pi^2_t(i)
L_{t+1}(O,\hat{\pi}^1_tP,\pi^1_tP,P_i,\pi^{3:N}_tP)\nonumber\\
\geq & \beta\sum_{i=1}^{L} \pi^2_t(i)L_{t+1}(O,\hat{\pi}^1_tP,\pi^1_tP,P_i,\pi^{3:N}_tP)\nonumber\\
= &\beta L_{t+1}(O,\hat{\pi}^1_tP,\pi^{1:N}_tP).
   \label{pfP:funrotate:eqorder}
\end{align}
The equalities in (\ref{pfP:funrotate:eqorder}) are true because of the linearity of $L_t$ (Lemma \ref{lm:funlin}).
The inequalities in (\ref{pfP:funrotate:eqorder}) are true because of (\ref{pfP:funrotate:eq2}).
\\
Combing (\ref{pfP:funrotate:eq1notend}) and (\ref{pfP:funrotate:eqorder}) we get
\begin{align}
&\beta L_{t+1}(SO,\hat{\pi}^1_tP,\pi^{1:N}_tP) \geq L_t(O,\hat{\pi}^1_t,\pi^{1:N}_t). 
\label{pfP:funrotate:eqorder1}\\
& L_t(O,\hat{\pi}^1_t,\pi^{1:N}_t) \geq \beta L_{t+1}(O,\hat{\pi}^1_tP,\pi^{1:N}_tP).
\label{pfP:funrotate:eqorder2}
\end{align}
When $O(N) = 1$,
\begin{align}
&L_t(O_t,\hat{\pi}^1_t,\pi^{1:N}_t)\nonumber\\
:= &V_t(O_t,\hat{\pi}^1_t,\pi^{2:N}_t)
-V_t(O_t,\pi^1_t,\pi^{2:N}_t)\nonumber\\
= &(\hat{\pi}^{1}_tR-\pi^{1}_tR) 
+\beta\sum_{i<L} (\hat{\pi}^{1}_t(i)-\pi^1_t(i))
 V_{t+1}(SO_t,P_i,\pi^{2:N}_tP)+\beta\sum_{i\geq L} (\hat{\pi}^{1}_t(i)-\pi^1_t(i))
 V_{t+1}(O_t,P_i,\pi^{2:N}_tP)\nonumber\\
= &(\hat{\pi}^{1}_t-\pi^{1}_t) R
+\beta\sum_{i=1}^{K} (\underline{\hat{\pi}}^{1}_t(i)-\underline{\pi}^1_t(i))
 V_{t+1}(SO_t,P_i,\pi^{2:N}_tP) 
+\beta\sum_{i=1}^{K} (\bar{\hat{\pi}}^{1}_t(i)-\bar{\pi}^1_t(i))
 V_{t+1}(O_t,P_i,\pi^{2:N}_tP)\nonumber\\
&+ \beta(V_{t+1}(O_t,P_L,\pi^{2:N}_tP)-V_{t+1}(SO_t,P_{L-1},\pi^{2:N}_tP) )\sum_{i=L}^{K}(\hat{\pi}^{1}_t(i)-\pi^1_t(i))\nonumber\\
=&(\hat{\pi}^{1}_t-\pi^{1}_t) R
+\beta L_{t+1}(SO_t,\underline{\hat{\pi}}^{1}_tP,\underline{\pi}^{1}_tP,\pi^{2:N}_tP) 
+\beta L_{t+1}(O_t,\bar{\hat{\pi}}^{1}_tP,\bar{\pi}^{1}_tP,\pi^{2:N}_tP)\nonumber\\
&+ \beta(V_{t+1}(O_t,P_L,\pi^{2:N}_tP)-V_{t+1}(SO_t,P_{L-1},\pi^{2:N}_tP) )\sum_{i=L}^{K}(\hat{\pi}^{1}_t(i)-\pi^1_t(i)).
\label{pfP:funrotate:eq1isend}
\end{align}
The second equality in (\ref{pfP:funrotate:eq1isend}) follows from the recursive equation for $V_t$ (eq. (\ref{eqVt})).
The third equality in (\ref{pfP:funrotate:eq1isend}) is true because of the definition
of $\underline{\pi}, \bar{\pi}$ given by (\ref{pfP:undxdef}) and (\ref{pfP:barxdef}).
The last equality in (\ref{pfP:funrotate:eq1isend}) follows from the linearity of $L_t$ (Lemma \ref{lm:funlin}).
\\
Furthermore, using (\ref{pfP:funrotate:eq1isend}) we get
\begin{align}
&L_t(O,\hat{\pi}^1_t,\pi^{1:N}_t) -\beta L_{t+1}(SO,\hat{\pi}^1_tP,\pi^{1:N}_tP)\nonumber\\
=&(\hat{\pi}^{1}_t-\pi^{1}_t) R
+\beta L_{t+1}(SO,\underline{\hat{\pi}}^{1}_tP,\underline{\pi}^{1}_tP,\pi^{2:N}_tP) 
+\beta L_{t+1}(O,\bar{\hat{\pi}}^{1}_tP,\bar{\pi}^{1}_tP,\pi^{2:N}_tP)\nonumber\\
&+ \beta(V_{t+1}(O,P_L,\pi^{2:N}_tP)-V_{t+1}(SO,P_{L-1},\pi^{2:N}_tP) )\sum_{i=L}^{K}(\hat{\pi}^{1}_t(i)-\pi^1_t(i))
\nonumber\\
& -\beta L_{t+1}(SO,\hat{\pi}^1_tP,\pi^1_tP,\pi^{2:N}_tP)\nonumber\\
=&(\hat{\pi}^{1}_t-\pi^{1}_t) R
+\beta L_{t+1}(O,\bar{\hat{\pi}}^{1}_tP,\bar{\pi}^{1}_tP,\pi^{2:N}_tP)
-\beta L_{t+1}(SO,\bar{\hat{\pi}}^{1}_tP,\bar{\pi}^{1}_tP,\pi^{2:N}_tP)\nonumber\\
&+ \beta(V_{t+1}(O,P_L,\pi^{2:N}_tP)-V_{t+1}(SO,P_{L},\pi^{2:N}_tP)\sum_{i=L}^{K}(\hat{\pi}^{1}_t(i)-\pi^1_t(i))\nonumber\\
\leq &(\hat{\pi}^{1}_t-\pi^{1}_t) R
+\beta (\bar{\hat{\pi}}^{1}_t-\bar{\pi}^{1}_t)PU\nonumber\\
&+ \beta(V_{t+1}(O,P_L,\pi^{2:N}_tP)-V_{t+1}(SO,P_{L},\pi^{2:N}_tP)\sum_{i=L}^{K}(\hat{\pi}^{1}_t(i)-\pi^1_t(i)).
					   	\label{pfP:funrotate:eqL0}
\end{align}
The first equality in (\ref{pfP:funrotate:eqL0}) follows from (\ref{pfP:funrotate:eq1isend}).
The second equality in (\ref{pfP:funrotate:eqL0}) follows from (\ref{pfP:underbarx}) and the linearity of $L_t$ (Lemma \ref{lm:funlin}).
The inequality in (\ref{pfP:funrotate:eqL0}) follows from the induction hypothesis for the upper bound of Property \ref{P:funrotate} at $t+1$ and the fact that $\bar{\hat{\pi}}^{1}_tP \geq_{st}\bar{\pi}^{1}_tP$.
\\
For the last term in (\ref{pfP:funrotate:eqL0}), because 
$\bar{\hat{\pi}}^{1}_t\geq_{st} \bar{\pi}^{1}_t$, we have
\begin{align}
\sum_{i=L}^{K}(\hat{\pi}^{1}_t(i)-\pi^1_t(i)) \geq 0 .\label{pfP:funrotate:eqL02}
\end{align} Moreover,
\begin{align}
&V_{t+1}(O,P_L,\pi^{2:N}_tP)-V_{t+1}(SO,P_{L},\pi^{2:N}_tP)\nonumber\\
=&L_{t+1}(O,P_L,P_{L-1},\pi^{2:N}_tP)-L_{t+1}(SO,P_L,P_{L-1},\pi^{2:N}_tP)\nonumber\\
&+V_{t+1}(O,P_{L-1},\pi^{2:N}_tP)-V_{t+1}(SO,P_{L-1},\pi^{2:N}_tP)\nonumber\\
=&L_{t+1}(O,P_L,P_{L-1},\pi^{2:N}_tP)-L_{t+1}(SO,P_L,P_{L-1},\pi^{2:N}_tP)\nonumber\\
&+V_{t+1}(O,P_{L-1},\pi^{2:N}_tP)-V_{t+1}(W_{12}...W_{(N-1)(N-2)}W_{N(N-1)}O,P_{L-1},\pi^{2:N}_tP)\nonumber\\
\leq&L_{t+1}(O,P_L,P_{L-1},\pi^{2:N}_tP)-L_{t+1}(SO,P_L,P_{L-1},\pi^{2:N}_tP)\nonumber\\
\leq & (P_L-P_{L-1})U .
\label{pfP:funrotate:eqL03}
\end{align}
The first equality in (\ref{pfP:funrotate:eqL03}) follows from the definition of $L_{t+1}$.
The second equality in (\ref{pfP:funrotate:eqL03}) is true because \\
$SO=W_{12}...W_{(N-1)(N-2)}W_{N(N-1)}O$.
The first inequality in (\ref{pfP:funrotate:eqL03}) follows by repeatedly using Property \ref{P:funswitch2} at $t+1$ and the fact that $\pi^{m}_tP \geq_{st} P_{L-1}$ for all $m=2,3,...,N$.
The second inequality in (\ref{pfP:funrotate:eqL03}) follows from the induction hypothesis for the upper bound of Property \ref{P:funrotate} at $t+1$ and the fact that $P_L\geq_{st} P_{L-1}$.
\\
Therefore, using (\ref{pfP:funrotate:eqL02}) and (\ref{pfP:funrotate:eqL03}) in (\ref{pfP:funrotate:eqL0}) give
\begin{align}
&L_t(O,\hat{\pi}^1_t,\pi^1_t,\pi^{2:N}_t) -\beta L_{t+1}(SO,\hat{\pi}^1_tP,\pi^1_tP,\pi^{2:N}_tP)\nonumber\\
\leq &(\hat{\pi}^{1}_t-\pi^{1}_t) R+\beta (\bar{\hat{\pi}}^{1}_t-\bar{\pi}^{1}_t)PU +\beta(P_L-P_{L-1})U \sum_{i=L}^{K}(\hat{\pi}^{1}_t(i)-\pi^1_t(i))\nonumber\\
= & (\hat{\pi}^{1}_t-\pi^{1}_t)R + \beta \sum_{i \geq L}(\hat{\pi}^{1}_t(i)-\pi^1_t(i))P_iU + \beta\sum_{i <L} (\hat{\pi}^{1}_t(i)-\pi^1_t(i))P_{L-1}U \nonumber\\
= & (\hat{\pi}^{1}_t-\pi^{1}_t)U.
       \label{pfP:funrotate:eqU}
\end{align}
The inequality in (\ref{pfP:funrotate:eqU}) follows from (\ref{pfP:funrotate:eqL0}), (\ref{pfP:funrotate:eqL02}) and (\ref{pfP:funrotate:eqL03}).
The first equality in (\ref{pfP:funrotate:eqU}) follows from the definition of $\bar{\hat{\pi}}^{1}_t$ and $\bar{\pi}^{1}_t$ given by (\ref{pfP:barxdef}).
The last equality in (\ref{pfP:funrotate:eqU}) follows from the definition of $U$.
\\\\
\textbf{Induction step for Property \ref{P:funrotate}:\\}
We first consider the lower bound of Property \ref{P:funrotate}. We want to show that
\begin{align}
L_t(O_t,\hat{\pi}^1_t,\pi^{1:N}_t) \geq L_t(S^{-m}O_t,\hat{\pi}^1_t,\pi^{1:N}_t).
\end{align}
(i) When $O_t(N)\neq 1$ (i.e. $n \neq N$), we also have $S^{-m}O_t(N) = O_t(m) \neq 1$.
Then,
\begin{align}
L_t(O_t,\hat{\pi}^1_t,\pi^{1:N}_t) 
\geq & \beta L_{t+1}(O_t,\hat{\pi}^1_tP,\pi^{1:N}_tP)\nonumber\\
=&\beta L_{t+1}(S^{m}S^{-m}O_t,\hat{\pi}^1_tP,\pi^{1:N}_tP)\nonumber\\
\geq&\beta L_{t+1}(S^{1-m}O_t,\hat{\pi}^1_tP,\pi^{1:N}_tP)\nonumber\\
\geq& L_t(S^{-m}O_t,\hat{\pi}^1_t,\pi^{1:N}_t).
\label{pfP:funrotate:eqlbi}
\end{align}
The first inequality in (\ref{pfP:funrotate:eqlbi}) follows from (\ref{pfP:funrotate:eqorder1}) and the fact that $O_t(N)\neq 1$.
The second inequality in (\ref{pfP:funrotate:eqlbi}) follows from the induction hypothesis for Property \ref{P:funrotate} at $t+1$.
The last inequality in (\ref{pfP:funrotate:eqlbi}) follows from (\ref{pfP:funrotate:eqorder2}) and the fact that $S^{-m}O_t(N)\neq 1$.
\\
This completes the proof of the lower bound of Property \ref{P:funrotate} for case (i).
\\\\
(ii) When $O_t(N)= 1$ (i.e. $n = N$).\\
Since $S^{-m}O_t(N) = O_t(m) \neq 1$, we get
\begin{align}
&L_t(S^{-m}O_t,\hat{\pi}^1_t,\pi^{1:N}_t)\nonumber\\
\leq 
&\beta L_{t+1}(S^{1-m}O_t,\hat{\pi}^1_tP,\pi^{1:N}_tP)\nonumber\\
=&\beta L_{t+1}(S^{1-m}O_t,\underline{\hat{\pi}}^1_tP,\underline{\pi}^1_tP,\pi^{2:N}_tP)
+\beta L_{t+1}(S^{1-m}O_t,\bar{\hat{\pi}}^1_tP,\bar{\pi}^1_tP,\pi^{2:N}_tP)\nonumber\\
&+\beta \sum_{i=L}^K(\hat{\pi}^1_t(i)-\pi^1_t(i))L_{t+1}(S^{1-m}O_t,P_L,P_{L-1},\pi^{2:N}_tP)
\label{pfP:funrotate:eqLSOt}
\end{align}
The inequality in (\ref{pfP:funrotate:eqLSOt}) follows from (\ref{pfP:funrotate:eqorder2}) and the fact that $S^{-m}O_t(N)\neq 1$.
The equality in (\ref{pfP:funrotate:eqLSOt}) follows from (\ref{pfP:underbarx}) and the linearity of $L_t$ (Lemma \ref{lm:funlin}).
\\
Since $O_t(N)=1 $, applying (\ref{pfP:funrotate:eq1isend}) we obtain
\begin{align}
& L_t(O_t,\hat{\pi}^1_t,\pi^{1:N}_t)-L_t(S^{-m}O_t,\hat{\pi}^1_t,\pi^{1:N}_t)\nonumber\\
=&(\hat{\pi}^{1}_t-\pi^{1}_t) R
+\beta L_{t+1}(SO_t,\underline{\hat{\pi}}^{1}_tP,\underline{\pi}^{1}_tP,\pi^{2:N}_tP) 
+\beta L_{t+1}(O_t,\bar{\hat{\pi}}^{1}_tP,\bar{\pi}^{1}_tP,\pi^{2:N}_tP)\nonumber\\
&+ \beta(V_{t+1}(O_t,P_L,\pi^{2:N}_tP)-V_{t+1}(SO_t,P_{L-1},\pi^{2:N}_tP) )\sum_{i=L}^{K}(\hat{\pi}^{1}_t(i)-\pi^1_t(i))
-L_t(S^{-m}O_t,\hat{\pi}^1_t,\pi^1_t,\pi^{2:N}_t)\nonumber\\
\geq &(\hat{\pi}^{1}_t-\pi^{1}_t) R
+\beta L_{t+1}(SO_t,\underline{\hat{\pi}}^{1}_tP,\underline{\pi}^{1}_tP,\pi^{2:N}_tP) 
-\beta L_{t+1}(S^{1-m}O_t,\underline{\hat{\pi}}^{1}_tP,\underline{\pi}^{1}_tP,\pi^{2:N}_tP) \nonumber\\
&+\beta L_{t+1}(O_t,\bar{\hat{\pi}}^{1}_tP,\bar{\pi}^{1}_tP,\pi^{2:N}_tP)
-\beta L_{t+1}(S^{1-m}O_t,\bar{\hat{\pi}}^{1}_tP,\bar{\pi}^{1}_tP,\pi^{2:N}_tP)\nonumber\\
&+ \beta(V_{t+1}(O_t,P_L,\pi^{2:N}_tP)-V_{t+1}(SO_t,P_{L-1},\pi^{2:N}_tP)-L_{t+1}(S^{1-m}O_t,P_L,P_{L-1},\pi^{2:N}_tP) )
 \sum_{i=L}^{K}(\hat{\pi}^{1}_t(i)-\pi^1_t(i))\nonumber\\
\geq &(\hat{\pi}^{1}_t-\pi^{1}_t) R
+\beta L_{t+1}(SO_t,\underline{\hat{\pi}}^{1}_tP,\underline{\pi}^{1}_tP,\pi^{2:N}_tP) 
-\beta L_{t+1}(S^{1-m}O_t,\underline{\hat{\pi}}^{1}_tP,\underline{\pi}^{1}_tP,\pi^{2:N}_tP) \nonumber\\
&+ \beta(V_{t+1}(O_t,P_L,\pi^{2:N}_tP)-V_{t+1}(SO_t,P_{L-1},\pi^{2:N}_tP)-L_{t+1}(S^{1-m}O_t,P_L,P_{L-1},\pi^{2:N}_tP) )
 \sum_{i=L}^{K}(\hat{\pi}^{1}_t(i)-\pi^1_t(i)).
\label{pfP:funrotate:eq1iN}
\end{align}
The equality in (\ref{pfP:funrotate:eq1iN}) follows from (\ref{pfP:funrotate:eq1isend}) and the fact that $O_t(N)= 1$.
The first inequality in (\ref{pfP:funrotate:eq1iN}) follows from (\ref{pfP:funrotate:eqLSOt}).
The second inequality in (\ref{pfP:funrotate:eq1iN}) follows from the induction hypothesis for the lower bound of Property \ref{P:funrotate} at $t+1$ and the fact that $\bar{\hat{\pi}}^{1}_tP\geq_{st}\bar{\pi}^1_tP$.
\\
Letting $\underline{O}_{t+1}:=S^{1-m}O_t$ and $\underline{n}:=N+1-m,\underline{m}:=N-m$, we have $\underline{m}<\underline{n}$ and
\begin{align}
&\underline{O}_{t+1}(\underline{n})=S^{1-m}O_t(\underline{n})=1,\\
&SO_t = S^{-(\underline{m})}\underline{O}_{t+1}.
\end{align}
Consequently, the induction hypothesis for the upper bound of Property \ref{P:funrotate} at $t+1$ gives
\begin{align}
& L_{t+1}(S^{1-m}O_t,\underline{\hat{\pi}}^{1}_tP,\underline{\pi}^{1}_tP,\pi^{2:N}_tP) 
-L_{t+1}(SO_t,\underline{\hat{\pi}}^1_tP,\underline{\pi}^1_tP,\pi^{2:N}_tP)\nonumber\\
=& L_{t+1}(\underline{O}_{t+1},\underline{\hat{\pi}}^{1}_tP,\underline{\pi}^{1}_tP,\pi^{2:N}_tP) 
-L_{t+1}(S^{-(\underline{m})}\underline{O}_{t+1},\underline{\hat{\pi}}^1_tP,\underline{\pi}^1_tP,\pi^{2:N}_tP)
\leq  (\underline{\hat{\pi}}^1_tP-\underline{\pi}^1_tP)U.
\label{pfP:funrotate:equndU}
\end{align}
Letting $\underline{m}':=1$, we have $\underline{m}'<n=N$ and
\begin{align}
&A_{\underline{m}'n}O_t = SO_t. \label{pfP:funrotate:eqAmn}
\end{align}
Therefore,
\begin{align}
&V_{t+1}(SO_t,P_{L-1},\pi^{2:N}_tP)-V_{t+1}(O_t,P_L,\pi^{2:N}_tP)
+L_{t+1}(S^{1-m}O_t,P_L,P_{L-1},\pi^{2:N}_tP) \nonumber\\
\leq 
&V_{t+1}(SO_t,P_{L-1},\pi^{2:N}_tP)-V_{t+1}(O_t,P_L,\pi^{2:N}_tP)
+L_{t+1}(O_t,P_L,P_{L-1},\pi^{2:N}_tP) \nonumber\\
=&V_{t+1}(A_{\underline{m}'n}O_t,P_{L-1},\pi^{2:N}_tP)-V_{t+1}(O_t,P_{L-1},\pi^{2:N}_tP)\nonumber\\
\leq& h-P_{L-1}R.
\label{pfP:funrotate:eqh}
\end{align}
The first inequality in (\ref{pfP:funrotate:eqh}) follows from the induction hypothesis for the lower bound of Property \ref{P:funrotate} at $t+1$ and the fact that $P_L\geq_{st} P_{L-1}$.
The equality in (\ref{pfP:funrotate:eqh}) follows from the definition of $L_{t+1}$ and (\ref{pfP:funrotate:eqAmn}).
The last inequality in (\ref{pfP:funrotate:eqh}) follows from the induction hypothesis for Property \ref{P:funlift} at $t+1$ and the fact that $P_{L-1}\in \pi P$, therefore $P_{L-1}\leq_{st}P_{L-1}P$ by Property \ref{P:PMForder}.
\\
Using (\ref{pfP:funrotate:equndU}) and (\ref{pfP:funrotate:eqh}) in (\ref{pfP:funrotate:eq1iN}) we obtain
\begin{align}
& L_t(O_t,\hat{\pi}^1_t,\pi^{1:N}_t)-L_t(S^{-m}O_t,\hat{\pi}^1_t,\pi^{1:N}_t)\nonumber\\
\geq &(\hat{\pi}^{1}_t-\pi^{1}_t) R
-\beta(\underline{\hat{\pi}}^1_tP-\underline{\pi}^1_tP)U
- \beta\sum_{i=L}^{K}(\hat{\pi}^{1}_t(i)-\pi^1_t(i))(h-P_{L-1}R)\nonumber\\
= &(\underline{\hat{\pi}}^1_t-\underline{\pi}^1_t) R+(\bar{\hat{\pi}}^1_t-\bar{\pi}^1_t) R
+\sum_{i=L}^{K}(\hat{\pi}^{1}_t(i)-\pi^1_t(i))(R_L-R_{L-1})\nonumber\\
&-\beta(\underline{\hat{\pi}}^1_tP-\underline{\pi}^1_tP)U
- \beta\sum_{i=L}^{K}(\hat{\pi}^{1}_t(i)-\pi^1_t(i))(h-P_{L-1}R)\nonumber\\
= &(\underline{\hat{\pi}}^1_t-\underline{\pi}^1_t) (R-\beta U)+(\bar{\hat{\pi}}^1_t-\bar{\pi}^1_t) R
+\sum_{i=L}^{K}(\hat{\pi}^{1}_t(i)-\pi^1_t(i))(R_L-R_{L-1}-\beta(h-P_{L-1}R))\nonumber\\
\geq &0.
\label{pfP:funrotate:eqLB}
\end{align}
The first inequality in (\ref{pfP:funrotate:eqLB}) follows from eqs (\ref{pfP:funrotate:equndU}) and (\ref{pfP:funrotate:eqh}) and the fact that $\sum_{i=L}^{K}(\hat{\pi}^{1}_t(i)-\pi^1_t(i))\geq 0$ (since $\hat{\pi}^{1}_t(i)\geq_{st}\pi^1_t$).
The first equality in (\ref{pfP:funrotate:eqLB}) follows from (\ref{pfP:underbarx}).
The last inequality in (\ref{pfP:funrotate:eqLB}) is true because:
the terms $(\underline{\hat{\pi}}^1_t-\underline{\pi}^1_t) (R-\beta U)$ and $(\bar{\hat{\pi}}^1_t-\bar{\pi}^1_t) R$ are positive by parts (iv) and (ii) of Property \ref{P:instreward} and the fact that $\underline{\hat{\pi}}^1_t\geq_{st} \underline{\pi}^1_t$ and $\bar{\hat{\pi}}^1_t\geq_{st}\bar{\pi}^1_t$;
the term $(R_L-R_{L-1}-\beta(h-P_{L-1}R))$ is positive by condition (A\ref{A:reward}).\\
The proof of the lower bound of Property \ref{P:funrotate} is now complete.
\\\\
Now consider the upper bound of Property \ref{P:funrotate}.
We want to show that
\begin{align}
L_t(O_t,\hat{\pi}^1_t,\pi^{1:N}_t) -L_t(S^{-m}O_t,\hat{\pi}^1_t,\pi^{1:N}_t) \leq (\hat{\pi}^1_t-\pi^1_t)U.
\end{align}
Let $O_t':=S^{N-n}O_t$;, then $O_t'(N)=1$ and $SO_t'(1)=1$.
Consequently, 
\begin{align}
L_t(O_t,\hat{\pi}^1_t,\pi^{1:N}_t) -L_t(S^{-m}O_t,\hat{\pi}^1_t,\pi^{1:N}_t)
\leq &
L_t(O_t',\hat{\pi}^1_t,\pi^{1:N}_t) -L_t(SO_t',\hat{\pi}^1_t,\pi^{1:N}_t) \nonumber\\
\leq &
L_t(O_t',\hat{\pi}^1_t,\pi^{1:N}_t) -\beta L_{t+1}(SO_t',\hat{\pi}^1_tP,\pi^{1:N}_tP) \nonumber\\
\leq &(\hat{\pi}^1_t-\pi^1_t)U .
\label{pfP:funrotate:eqUB}
\end{align}
The first inequality in (\ref{pfP:funrotate:eqUB}) is true because of the lower bound of Property \ref{P:funrotate} at $t$.
The second inequality in (\ref{pfP:funrotate:eqUB}) follows from (\ref{pfP:funrotate:eqorder2}) and the fact that 
$SO_t'(N)\neq 1$.
The third inequality in (\ref{pfP:funrotate:eqUB}) follows from (\ref{pfP:funrotate:eqU}) and the fact that 
$O_t'(N)= 1$.
\\
This completes the proof of Property \ref{P:funrotate} at time $t$.
\\\\
\textbf{Induction step for Property \ref{P:funswitch}:\\} \label{pfP:funswitch}
(i) When $O_t(N)\neq 1$ (i.e. $n \neq N$), assume $O_t(N)=2$ without loss of generality. Then because of (\ref{pfP:funrotate:eq1notend}),  
\begin{align}
&L_t(O_t,\hat{\pi}^1_t,\pi^{1:N}_t) -L_t(W_{nm}O_t,\hat{\pi}^1_t,\pi^{1:N}_t)\nonumber\\
=& \beta\sum_{i<L} \pi^{2}(i)(L_{t+1}(SO_t,\hat{\pi}^1_tP,\pi^1_tP,P_i,\pi^{3:N}_tP) -L_{t+1}(S(W_{nm}O_t),\hat{\pi}^1_tP,\pi^1_tP,P_i,\pi^{3:N}_tP))\nonumber\\
& + \beta\sum_{i\geq L} \pi^{2}(i)(L_{t+1}(O_t,\hat{\pi}^1_tP,\pi^1_tP,P_i,\pi^{3:N}_tP) -L_{t+1}(W_{nm}O_t,\hat{\pi}^1_tP,\pi^1_tP,P_i,\pi^{3:N}_tP))\nonumber\\
=& \beta\sum_{i<L} \pi^{2}(i)(L_{t+1}(SO_t,\hat{\pi}^1_tP,\pi^1_tP,P_i,\pi^{3:N}_tP) -L_{t+1}(W_{(n+1)(m+1)}(SO_t),\hat{\pi}^1_tP,\pi^1_tP,P_i,\pi^{3:N}_tP))\nonumber\\
& + \beta\sum_{i\geq L} \pi^{2}(i)(L_{t+1}(O_t,\hat{\pi}^1_tP,\pi^1_tP,P_i,\pi^{3:N}_tP) -L_{t+1}(W_{nm}O_t,\hat{\pi}^1_tP,\pi^1_tP,P_i,\pi^{3:N}_tP)).
\label{pfP:funswitch:eq1}
\end{align}
The first equality in (\ref{pfP:funswitch:eq1}) follows from (\ref{pfP:funrotate:eq1notend}). The second equality is true because
$S(W_{nm}O_t) = W_{(n+1)(m+1)} (SO_t)$.
\\
By the induction hypothesis for Property \ref{P:funswitch}, each term in (\ref{pfP:funswitch:eq1}) is positive and smaller than
$(\hat{\pi}^1_tP-\pi^1_tP)M $. Thus, 
\begin{align}
0\leq &L_t(O_t,\hat{\pi}^1_t,\pi^{1:N}_t) -L_t(W_{nm}O_t,\hat{\pi}^1_t,\pi^{1:N}_t)\nonumber\\
=& \beta\sum_{i<L} \pi^{2}(i)(L_{t+1}(SO_t,\hat{\pi}^1_tP,\pi^1_tP,P_i,\pi^{3:N}_tP) -L_{t+1}(W_{(n+1)(m+1)}(SO_t),\hat{\pi}^1_tP,\pi^1_tP,P_i,\pi^{3:N}_tP))\nonumber\\
& + \beta\sum_{i\geq L} \pi^{2}(i)(L_{t+1}(O_t,\hat{\pi}^1_tP,\pi^1_tP,P_i,\pi^{3:N}_tP) -L_{t+1}(W_{nm}O_t,\hat{\pi}^1_tP,\pi^1_tP,P_i,\pi^{3:N}_tP))\nonumber\\
\leq &\beta(\hat{\pi}^1_tP-\pi^1_tP)M \nonumber\\
\leq &(\hat{\pi}^1_t-\pi^1_t)M.
\label{pfP:funswitch:eq2}
\end{align}
The first and second inequalities in (\ref{pfP:funswitch:eq2}) follow from the induction hypothesis for Property \ref{P:funswitch}.
The equality in (\ref{pfP:funswitch:eq2}) follow from (\ref{pfP:funswitch:eq1}).
The last inequality in (\ref{pfP:funswitch:eq2}) holds by part (iii) of Property \ref{P:instreward} and the fact that 
$\hat{\pi}^1_t\geq_{st}\pi^1_t$.
\\
The proof of Property \ref{P:funswitch} is now complete when $O_t(N)\neq 1$.
\\\\
(ii) $O_t(N)= 1$ (i.e. $n = N$).\\
We first consider the lower-bound. We want to show that
\begin{align}
L_t(O_t,\hat{\pi}^1_t,\pi^{1:N}_t) -L_t(W_{Nm}O_t,\hat{\pi}^1_t,\pi^{1:N}_t) \geq 0.
\end{align}
Using (\ref{pfP:underbarx}) and the linearity of $L_t$ (Lemma \ref{lm:funlin}) we get
\begin{align}
&L_t(O_t,\hat{\pi}^1_t,\pi^{1:N}_t) -L_t(W_{Nm}O_t,\hat{\pi}^1_t,\pi^{1:N}_t)\nonumber\\
=&L_t(O_t,\underline{\hat{\pi}}^1_t,\underline{\pi}^1_t,\pi^{2:N}_t) -L_t(W_{Nm}O_t,\underline{\hat{\pi}}^1_t,\underline{\pi}^1_t,\pi^{2:N}_t)\nonumber\\
&+L_t(O_t,\bar{\hat{\pi}}^1_t,\hat{\pi}^1_t,\pi^{2:N}_t) -L_t(W_{Nm}O_t,\bar{\hat{\pi}}^1_t,\hat{\pi}^1_t,\pi^{2:N}_t)\nonumber\\
&+ \left[\sum_{i=L}^K(\hat{\pi}^1_t(i)-\pi^1_t(i))\right][L_t(O_t,e_L,e_{L-1},\pi^{2:N}_t) -L_t(W_{Nm}O_t,e_L,e_{L-1},\pi^{2:N}_t)].
\label{pfP:funswitch:I0}
\end{align}
We consider each of the terms
\begin{enumerate}[(a)]
\item $L_t(O_t,\underline{\hat{\pi}}^1_t,\underline{\pi}^1_t,\pi^{2:N}_t) -L_t(W_{Nm}O_t,\underline{\hat{\pi}}^1_t,\underline{\pi}^1_t,\pi^{2:N}_t)$.
\item
$L_t(O_t,\bar{\hat{\pi}}^1_t,\hat{\pi}^1_t,\pi^{2:N}_t) -L_t(W_{Nm}O_t,\bar{\hat{\pi}}^1_t,\hat{\pi}^1_t,\pi^{2:N}_t)$.
\item
$\left[\sum_{i=L}^K(\hat{\pi}^1_t(i)-\pi^1_t(i))\right][L_t(O_t,e_L,e_{L-1},\pi^{2:N}_t) -L_t(W_{Nm}O_t,e_L,e_{L-1},\pi^{2:N}_t)]$.
\end{enumerate}
that appear in the right hand side of (\ref{pfP:funswitch:I0}) separately.
We do this because the  channel orderings are different in each of the tree terms, different methods are needed to establish the bounds.
\\\\
(a) Consider the first term.\\ 
Let $O_t' =S(W_{Nm}O_t)=W_{1m+1}(SO_t)$, then $O_t'(m+1)=1$ and $W_{m+1,1}O_t' = SO_t$.
Therefore,
\begin{align}
&L_t(O_t,\underline{\hat{\pi}}^1_t,\underline{\pi}^1_t,\pi^{2:N}_t) -L_t(W_{Nm}O_t,\underline{\hat{\pi}}^1_t,\underline{\pi}^1_t,\pi^{2:N}_t)
\nonumber\\
= & (\underline{\hat{\pi}}^1_t-\underline{\pi}^1_t) R
+\beta L_{t+1}(SO_t,\underline{\hat{\pi}}^{1}_tP,\underline{\pi}^{1}_tP,\pi^{2:N}_tP) 
-L_t(W_{Nm}O_t,\underline{\hat{\pi}}^1_t,\underline{\pi}^1_t,\pi^{2:N}_t)
\nonumber\\
\geq & (\underline{\hat{\pi}}^1_t-\underline{\pi}^1_t) R
+\beta L_{t+1}(SO_t,\underline{\hat{\pi}}^{1}_tP,\underline{\pi}^{1}_tP,\pi^{2:N}_tP) 
-\beta L_{t+1}(S(W_{Nm}O_t),\underline{\hat{\pi}}^1_tP,\underline{\pi}^1_tP,\pi^{2:N}_tP)
\nonumber\\
= & (\underline{\hat{\pi}}^1_t-\underline{\pi}^1_t) R
-\beta (L_{t+1}(O_t',\underline{\hat{\pi}}^{1}_tP,\underline{\pi}^{1}_tP,\pi^{2:N}_tP) 
-L_{t+1}(W_{m+1,1}O_t',\underline{\hat{\pi}}^1_tP,\underline{\pi}^1_tP,\pi^{2:N}_tP))
\nonumber\\
\geq & (\underline{\hat{\pi}}^1_t-\underline{\pi}^1_t) R - \beta(\underline{\hat{\pi}}^1_tP-\underline{\pi}^1_tP)M
\nonumber\\ \geq& 0.
\label{pfP:funswitch:equnderlineLB}
\end{align}
The first equality in (\ref{pfP:funswitch:equnderlineLB}) follows from (\ref{pfP:funrotate:eq1isend}), the fact that $O_t(N)=1$ and that fact that $\underline{\hat{\pi}}^1_t(i)=\underline{\pi}^1_t(i) =0$ for $i\geq L$.
The first inequality in (\ref{pfP:funswitch:equnderlineLB}) follows from (\ref{pfP:funrotate:eqorder1}) and that fact that $W_{Nm}O_t(N)\neq 1$.
The second inequality in (\ref{pfP:funswitch:equnderlineLB}) follows from the induction hypothesis for the upper bound of Property \ref{P:funswitch} at $t+1$ and the fact that $\underline{\hat{\pi}}^1_tP\geq_{st}\underline{\pi}^1_tP$ (since $\underline{\hat{\pi}}^1_t\geq_{st}\underline{\pi}^1_t$ and Property \ref{P:preserveorder}).
The last inequality in (\ref{pfP:funswitch:equnderlineLB}) holds by part (iv) of Property \ref{P:instreward}, the fact that $\underline{\hat{\pi}}^1_t\geq_{st}\underline{\pi}^1_t$ and that fact that $\underline{\hat{\pi}}^1_t(i)=\underline{\pi}^1_t(i) =0$ for $i\geq L$.
\\\\
(b) Consider the second term.\\ 
Similar to case (a), we have
\begin{align}
&L_t(O_t,\bar{\hat{\pi}}^1_t,\bar{\pi}^1_t,\pi^{2:N}_t) -L_t(W_{Nm}O_t,\bar{\hat{\pi}}^1_t,\bar{\pi}^1_t,\pi^{2:N}_t)
\nonumber\\
= & (\bar{\hat{\pi}}^1_t-\bar{\pi}^1_t) R
+\beta L_{t+1}(O_t,\bar{\hat{\pi}}^{1}_tP,\bar{\pi}^{1}_tP,\pi^{2:N}_tP) 
-L_t(W_{Nm}O_t,\bar{\hat{\pi}}^1_t,\bar{\pi}^1_t,\pi^{2:N}_t)
\nonumber\\
\geq & (\bar{\hat{\pi}}^1_t-\bar{\pi}^1_t) R
+\beta L_{t+1}(SO_t,\bar{\hat{\pi}}^{1}_tP,\bar{\pi}^{1}_tP,\pi^{2:N}_tP) 
-L_t(W_{Nm}O_t,\bar{\hat{\pi}}^1_t,\bar{\pi}^1_t,\pi^{2:N}_t)
\nonumber\\
\geq & (\bar{\hat{\pi}}^1_t-\bar{\pi}^1_t) R
+\beta L_{t+1}(SO_t,\bar{\hat{\pi}}^{1}_tP,\bar{\pi}^{1}_tP,\pi^{2:N}_tP) 
-\beta L_{t+1}(S(W_{Nm}O_t),\bar{\hat{\pi}}^1_tP,\bar{\pi}^1_tP,\pi^{2:N}_tP)
\nonumber\\
= & (\underline{\hat{\pi}}^1_t-\underline{\pi}^1_t) R
-\beta (L_{t+1}(O_t',\bar{\hat{\pi}}^{1}_tP,\bar{\pi}^{1}_tP,\pi^{2:N}_tP) 
-L_{t+1}(W_{m+1,1}O_t',\bar{\hat{\pi}}^1_tP,\bar{\pi}^1_tP,\pi^{2:N}_tP))
\nonumber\\
\geq & (\bar{\hat{\pi}}^1_t-\bar{\pi}^1_t) R - \beta(\bar{\hat{\pi}}^1_tP-\bar{\pi}^1_tP)M
\nonumber\\ \geq& 0.
\label{pfP:funswitch:eqbarLB}
\end{align}
The first equality in (\ref{pfP:funswitch:eqbarLB}) follows from (\ref{pfP:funrotate:eq1isend}), the fact that $O_t(N)=1$ and that fact that $\bar{\hat{\pi}}^1_t(i)=\bar{\pi}^1_t(i) =0$ for $i<L$.
The first inequality in (\ref{pfP:funswitch:eqbarLB}) follows from the induction hypothesis for the lower bound of Property \ref{P:funrotate} at $t+1$, the fact that $\bar{\hat{\pi}}^1_tP\geq_{st}\bar{\pi}^1_tP$ (since $\bar{\hat{\pi}}^1_t\geq_{st}\bar{\pi}^1_t$ and Property \ref{P:preserveorder})
and the fact that $SO_t=S^{-(N-1)}O_t$ and $O_t(N)=1$.
The second inequality in (\ref{pfP:funswitch:eqbarLB}) follows from (\ref{pfP:funrotate:eqorder1}) and that fact that $W_{Nm}O_t(N)\neq 1$.
The third inequality in (\ref{pfP:funswitch:eqbarLB}) follows from the induction hypothesis for the upper bound of Property \ref{P:funswitch} at $t+1$ and the fact that $\bar{\hat{\pi}}^1_tP\geq_{st}\bar{\pi}^1_tP$.
The last inequality in (\ref{pfP:funswitch:eqbarLB}) holds by part (iv) of Property \ref{P:instreward}, the fact that $\bar{\hat{\pi}}^1_tP\geq_{st}\bar{\pi}^1_tP$ and that fact that $\bar{\hat{\pi}}^1_t(i)=\bar{\pi}^1_t(i) =0$ for $i< L$.
\\\\
(c) Consider the third part.\\ 
Assume $O_t(m)=2$ without any loss of generality. Then $W_{Nm}O_t(N)=2$. Therefore, 
\begin{align}
&L_t(O_t,e_L,e_{L-1},\pi^{2:N}_t) -L_t(W_{Nm}O_t,e_L,e_{L-1},\pi^{2:N}_t) \nonumber\\
= & R_L - R_{L-1} +\beta [
  V_{t+1}(O_t,P_L,\pi^{2:N}_tP)
  -V_{t+1}(SO_t,P_{L-1},\pi^{2:N}_tP) ]\nonumber\\
  &-\beta\sum_{i<L} \pi^{2}_t(i)
 L_{t+1}(SW_{Nm}O_t,P_L,P_{L-1},P_i,\pi^{3:N}_tP)
 -\beta\sum_{i\geq L} \pi^{2}_t(i)
L_{t+1}(W_{Nm}O_t,P_L,P_{L-1},P_i,\pi^{3:N}_tP)\nonumber\\  	
= & R_L - R_{L-1} \nonumber\\
  & +\beta\sum_{i<L}\pi^2(i)[
   V_{t+1}(O_t,P_L,P_i,\pi^{3:N}_tP) -V_{t+1}(SO_t,P_{L-1},P_i,\pi^{3:N}_tP) \nonumber\\
   &\qquad -L_{t+1}(SW_{Nm}O_t,P_L,P_{L-1},P_i,\pi^{3:N}_tP) ]\nonumber\\
  &  +\beta\sum_{i\geq L}\pi^2(i)[
   V_{t+1}(O_t,P_L,P_i,\pi^{3:N}_tP) -V_{t+1}(SO_t,P_{L-1},P_i,\pi^{3:N}_tP) \nonumber\\
   &\qquad - L_{t+1}(W_{Nm}O_t,P_L,P_{L-1},P_i,\pi^{3:N}_tP) ].
\label{pfP:funswitch:eqKLB1}
\end{align}
The first equality in (\ref{pfP:funswitch:eqKLB1}) follows from (\ref{pfP:funrotate:eq1notend}) and (\ref{pfP:funrotate:eq1isend}).
The last equality in (\ref{pfP:funswitch:eqKLB1}) holds because of Lemma \ref{lm:funlin}.
\\
Let $O_t' :=S(W_{Nm}O_t)=W_{1m+1}(SO_t)$; then $O_t'(m+1)=1$ and $W_{m+1,1}O_t' = SO_t$.\\
For each term in the first sum in (\ref{pfP:funswitch:eqKLB1}), we have 
$P_{L-1} \geq_{st} P_i $ ($i<L$ in the first sum in (\ref{pfP:funswitch:eqKLB1})). Therefore,
\begin{align}
   & V_{t+1}(O_t,P_L,P_i,\pi^{3:N}_tP)-V_{t+1}(SO_t,P_{L-1},P_i,\pi^{3:N}_tP) 
   -L_{t+1}(SW_{Nm}O_t,P_L,P_{L-1},P_i,\pi^{3:N}_tP)\nonumber\\
 =  & V_{t+1}(O_t,P_L,P_i,\pi^{3:N}_tP)-V_{t+1}(W_{m+1,1}O_t',P_{L-1},P_i,\pi^{3:N}_tP) \nonumber\\
   & -V_{t+1}(O_t',P_L,P_i,\pi^{3:N}_tP)+V_{t+1}(O_t',P_{L-1},P_i,\pi^{3:N}_tP)\nonumber\\
 \geq  &V_{t+1}(O_t,P_L,P_i,\pi^{3:N}_tP)-V_{t+1}(O_t',P_L,P_i,\pi^{3:N}_tP).
\label{pfP:funswitch:eqKLBp11}
\end{align}
The equality in (\ref{pfP:funswitch:eqKLBp11}) follows from the definition of $L_{t+1}$.
The inequality in (\ref{pfP:funswitch:eqKLBp11}) follows from the induction hypothesis for the lower bound of Property \ref{P:funswitch2} at $t+1$ and the fact that $P_{L-1} \geq_{st} P_i $.
\\
Furthermore, since $P_L\geq_{st} \pi^{O_t(l)}_tP $ for all $l=1,2,...,N$ by Property \ref{P:PMForder}, repeatedly applying Property \ref{P:funswitch2} at $t+1$ we obtain
\begin{align}
   &V_{t+1}(O_t,P_L,P_i,\pi^{3:N}_tP)\nonumber\\
 \geq  &V_{t+1}(W_{(m+2)(m+1)}...W_{N(N-1)}O_t,P_L,P_i,\pi^{3:N}_tP)\nonumber\\
 =&V_{t+1}(A_{N(m+1)}O_t,P_L,P_i,\pi^{3:N}_tP),
\label{pfP:funswitch:eqKLBp12}
\end{align}
where $A_{Nm+1}$ is the operator defined by (\ref{eq:AOt}).
The equality in (\ref{pfP:funswitch:eqKLBp12}) is true because 
$W_{(m+2)(m+1)}...W_{N(N-1)}O_t=A_{N(m+1)}O_t$.
Note that 
\begin{align}
A_{m1}(A_{N(m+1)}O_t) =S(W_{Nm}O_t)=O_t', A_{N(m+1)}O_t(m)=O_t(m)=2.
\label{pfP:funswitch:eqA}
\end{align}
Consequently,
\begin{align}
   & V_{t+1}(O_t,P_L,P_i,\pi^{3:N}_tP)-V_{t+1}(SO_t,P_{L-1},P_i,\pi^{3:N}_tP) 
   -L_{t+1}(SW_{Nm}O_t,P_L,P_{L-1},P_i,\pi^{3:N}_tP)\nonumber\\
 \geq  &V_{t+1}(A_{N(m+1)}O_t,P_L,P_i,\pi^{3:N}_tP)-V_{t+1}(O_t',P_L,P_i,\pi^{3:N}_tP)\nonumber\\
 = &V_{t+1}(A_{N(m+)1}O_t,P_L,P_i,\pi^{3:N}_tP)-V_{t+1}(A_{m1}(A_{N(m+1)}O_t),P_L,P_i,\pi^{3:N}_tP) \nonumber\\
 \geq & -(h- P_{i}P^{N-m}R ).
\label{pfP:funswitch:eqKLBp1}
\end{align}
The first inequality in (\ref{pfP:funswitch:eqKLBp1}) follows from (\ref{pfP:funswitch:eqKLBp11}) and (\ref{pfP:funswitch:eqKLBp12}).
The equality in (\ref{pfP:funswitch:eqKLBp1}) follows from (\ref{pfP:funswitch:eqA}).
The second inequality in (\ref{pfP:funswitch:eqKLBp1}) follows from the induction hypothesis for Property \ref{P:funlift} at $t+1$ and the fact that $P_i\in \pi P$, therefore $P_i \leq_{st} P_{L-1} \leq_{st}P_iP $ for $i<L$ by Property \ref{P:PMForder}.
\\
For each term in the second sum in (\ref{pfP:funswitch:eqKLB1}), we have 
\begin{align}
   & V_{t+1}(O_t,P_L,P_i,\pi^{3:N}_tP)-V_{t+1}(SO_t,P_{L-1},P_i,\pi^{3:N}_tP) 
   -L_{t+1}(W_{Nm}O_t,P_L,P_{L-1},P_i,\pi^{3:N}_tP)\nonumber\\
 \geq & V_{t+1}(O_t,P_L,P_i,\pi^{3:N}_tP)-V_{t+1}(SO_t,P_{L-1},P_i,\pi^{3:N}_tP) 
   -L_{t+1}(O_t,P_L,P_{L-1},P_i,\pi^{3:N}_tP)\nonumber\\
 = &V_{t+1}(O_t,P_{L-1},P_i,\pi^{3:N}_tP)-V_{t+1}(SO_t,P_{L-1},P_i,\pi^{3:N}_tP)\nonumber\\
 = &V_{t+1}(O_t,P_{L-1},P_i,\pi^{3:N}_tP)-V_{t+1}(A_{N1}O_t,P_{L-1},P_i,\pi^{3:N}_tP)\nonumber\\
 \geq & -(h-P_{L-1}R).
\label{pfP:funswitch:eqKLBp2}
\end{align}
The first inequality in (\ref{pfP:funswitch:eqKLBp2}) follows from the induction hypothesis for the lower bound of Property \ref{P:funswitch} at $t+1$ and the fact that $P_L\geq_{st}P_{L-1}$.
The fist equality in (\ref{pfP:funswitch:eqKLBp2}) follows from the definition of $L_{t+1}$ (eq. (\ref{eq:L})).
The second equality in (\ref{pfP:funswitch:eqKLBp2}) follows from the fact that $SO_t=A_{N1}O_t$.
The last inequality in (\ref{pfP:funswitch:eqKLBp2}) follows from the induction hypothesis for Property \ref{P:funlift} at $t+1$ and the fact that $P_{L-1}\leq_{st}P_{L-1}P $.
\\
Using the lower bounds provided by (\ref{pfP:funswitch:eqKLBp1}) and (\ref{pfP:funswitch:eqKLBp2}) for terms in (\ref{pfP:funswitch:eqKLB1}), we obtain
\begin{align}
&L_t(O_t,e_L,e_{L-1},\pi^{2:N}_t) -L_t(W_{Nm}O_t,e_L,e_{L-1},\pi^{2:N}_t) \nonumber\\
\geq & R_L - R_{L-1} -\beta \sum_{i< L}\pi^2_t(i)(h-P_{i}P^{N-m}R)-\beta \sum_{i \geq L}\pi^2_t(i)(h-P_{L-1}R)  \nonumber\\
\geq & R_L - R_{L-1} -\beta (h- P_{L-1}R) \geq 0.
\label{pfP:funswitch:eqKLB3}
\end{align}
The first inequality in (\ref{pfP:funswitch:eqKLB3}) follows from (\ref{pfP:funswitch:eqKLBp1}) and (\ref{pfP:funswitch:eqKLBp2}).
The second inequality in (\ref{pfP:funswitch:eqKLB3}) follows from part (ii) of Property \ref{P:instreward} and the fact that
$P_{i}\in \pi P$, therefore $P_{i}P^{N-m} \in \pi P^2$, thus $P_{i}P^{N-m} \geq_{st} P_{L-1}$ by Property \ref{P:PMForder}. The last inequality in (\ref{pfP:funswitch:eqKLB3}) holds by condition (A\ref{A:reward}).
\\\\
Using the lower bounds given by (\ref{pfP:funswitch:equnderlineLB}), (\ref{pfP:funswitch:eqbarLB}) and 
(\ref{pfP:funswitch:eqKLB3}) for the three terms (a), (b) and (c), respectively, in \ref{pfP:funswitch:I0}, we obtain
\begin{align}
&L_t(O_t,\hat{\pi}^1_t,\pi^1_t,\pi^{2:N}_t) -L_t(W_{Nm}O_t,\hat{\pi}^1_t,\pi^1_t,\pi^{2:N}_t)\geq 0.
\end{align}
This completes the proof for the lower bound of Property \ref{P:funswitch} when $O_t(N)=1$ (case (ii)). 
\\\\
We now proceed to establish the upper bound of Property \ref{P:funswitch} when $O_t(N)=1$ (case (ii)). 
We want to show that
\begin{align}
L_t(O_t,\hat{\pi}^1_t,\pi^{1:N}_t) -L_t(W_{Nm}O_t,\hat{\pi}^1_t,\pi^{1:N}_t)  \leq (\hat{\pi}^1_t-\pi^1_t)M.
\end{align}
Assume $O_t(m)=2$ without any loss of generality; then $W_{Nm}O_t(N)=2$.
Therefore, 
\begin{align}
&L_t(O_t,\hat{\pi}^1_t,\pi^{1:N}_t) -L_t(W_{Nm}O_t,\hat{\pi}^1_t,\pi^{1:N}_t) \nonumber\\
= & L_t(O_t,\hat{\pi}^1_t,\pi^{1:N}_t)  - \beta L_{t+1}(SO_t,\hat{\pi}^1_tP,\pi^{1:N}_tP) \nonumber\\
&+\beta L_{t+1}(SO_t,\hat{\pi}^1_tP,\pi^{1:N}_tP) - L_t(W_{Nm}O_t,\hat{\pi}^1_t,\pi^{1:N}_t)\nonumber\\
\leq & (\hat{\pi}^1_t-\pi^1_t)U
+\beta L_{t+1}(SO_t,\hat{\pi}^1_tP,\pi^{1:N}_tP) - L_t(W_{Nm}O_t,\hat{\pi}^1_t,\pi^{1:N}_t)\nonumber\\
\leq & (\hat{\pi}^1_t-\pi^1_t)U
+\beta L_{t+1}(S(W_{Nm}O_t),\hat{\pi}^1_tP,\pi^{1:N}_tP) - L_t(W_{Nm}O_t,\hat{\pi}^1_t,\pi^{1:N}_t)\nonumber\\
= & (\hat{\pi}^1_t-\pi^1_t)U
+\beta L_{t+1}(S(W_{Nm}O_t),\hat{\pi}^1_tP,\pi^{1:N}_tP) 
- \beta\sum_{i<L} \pi^2_t(i) L_t(S(W_{Nm}O_t),\hat{\pi}^1_tP,\pi^{1}_tP,P_i,\pi^{3:N}_tP)\nonumber\\
&\qquad - \beta\sum_{i\geq L} \pi^2_t(i) L_{t+1}(W_{Nm}O_t,\hat{\pi}^1_tP,\pi^{1}_tP,P_i,\pi^{3:N}_tP)\nonumber\\
= & (\hat{\pi}^1_t-\pi^1_t)U
 + \beta\sum_{i\geq L} \pi^2_t(i) (
L_{t+1}(S(W_{Nm}O_t),\hat{\pi}^1_tP,\pi^{1}_tP,P_i,\pi^{3:N}_tP)-L_{t+1}(W_{Nm}O_t,\hat{\pi}^1_tP,\pi^{1}_tP,P_i,\pi^{3:N}_tP))\nonumber\\
\leq & (\hat{\pi}^1_t-\pi^1_t)U
 + \beta\sum_{i\geq L} \pi^2_t(i) (\hat{\pi}^1_tP-\pi^{1}_tP)U\nonumber\\
\leq & (\hat{\pi}^1_t-\pi^1_t)U
 + \beta\sum_{i\geq L} p_{Ki} (\hat{\pi}^1_tP-\pi^{1}_tP)U\nonumber\\
= &(\hat{\pi}^1_t-\pi^{1}_t)M.
\label{pfP:funswitch:eqI0n1}
\end{align}
The first inequality in (\ref{pfP:funswitch:eqI0n1}) follows from (\ref{pfP:funrotate:eqU}).
The second inequality in (\ref{pfP:funswitch:eqI0n1}) follows from the induction hypothesis for the lower bound of Property \ref{P:funswitch} at $t+1$, the fact that $SO_t=W_{(m+1),1}(S(W_{Nm}O_t))$ and the fact that $\hat{\pi}^1_t\geq_{st} \pi^{1}_t$.
The second equality in (\ref{pfP:funswitch:eqI0n1}) follows from (\ref{pfP:funrotate:eq1notend}).
The third equality in (\ref{pfP:funswitch:eqI0n1}) follows from the linearity of the function $L_t$ (Lemma \ref{lm:funlin}).
The third inequality in (\ref{pfP:funswitch:eqI0n1}) follows from the induction hypothesis for the upper bound of Property \ref{P:funrotate} and the fact that $\hat{\pi}^1_tP\geq_{st} \pi^{1}_tP$ (since $\hat{\pi}^1_t\geq_{st}\pi^1_t$ and Property \ref{P:preserveorder}).
The last inequality in (\ref{pfP:funswitch:eqI0n1}) is true because $\pi^2_t\leq_{st} P_K$.
The last equality in (\ref{pfP:funswitch:eqI0n1}) follows from the definition of $M$.
\\
The proof of the upper bound of Property \ref{P:funswitch} at $t$ is now complete.
The proof of the induction step for Property \ref{P:funswitch} at $t$ is also complete.
\\\\
\textbf{Induction step for Property \ref{P:funlift}:\\}
(i) When $O_t(N)\neq 1$ (i.e. $n \neq N$), assume $O_t(N)=N$ without loss of generality. Then, 
\begin{align}
&V_t(A_{nm}O_t,\pi^{1:N}_t) -V_t(O_t,\pi^{1:N}_t) \nonumber\\
= &\sum_{i<L}\pi^{N}_t(i)[V_{t+1}(S(A_{nm}O_t),\pi^{1:N-1}_tP,P_i)  -V_{t+1}(SO_t,\pi^{1:N-1}_tP,P_i)] \nonumber\\
   &+   \sum_{i \geq L}\pi^{N}_t(i)[V_{t+1}(A_{nm}O_t,\pi^{1:N-1}_tP,P_i)  -V_{t+1}(O_t,\pi^{1:N-1}_tP,P_i)] \nonumber\\
= &\sum_{i<L}\pi^{N}_t(i)[V_{t+1}(A_{(n+1),(m+1)}(SO_t),\pi^{1:N-1}_tP,P_i)  -V_{t+1}(SO_t,\pi^{1:N-1}_tP,P_i)] \nonumber\\
   &+   \sum_{i \geq L}\pi^{N}_t(i)[V_{t+1}(A_{nm}O_t,\pi^{1:N-1}_tP,P_i)  -V_{t+1}(O_t,\pi^{1:N-1}_tP,P_i)] \nonumber\\
\leq & \sum_{i<L}\pi^N_t(i)(h-\pi^1_tP(P^{N-n-1}R)) +   \sum_{i \geq L}\pi^N_t(i)(h-\pi^1_tP(P^{N-n}R)) \nonumber\\   
\leq & h-\pi^1_tP^{N-n}R.
\label{pfP:funlift:eqmg0}
\end{align}
The first equality in (\ref{pfP:funlift:eqmg0}) follows from the recursive equation for $V_t$ (eq. (\ref{eqVt})).
The second equality in (\ref{pfP:funlift:eqmg0}) is true because $S(A_{nm}O_t)=A_{(n+1),(m+1)}(SO_t)$.
The first inequality in (\ref{pfP:funlift:eqmg0}) follows from the induction hypothesis for Property \ref{P:funlift}
and the fact that $\pi^1_tP\leq_{st}\pi^1_tP^2$ (Property \ref{P:preserveorder}).
The last inequality in (\ref{pfP:funlift:eqmg0}) follows from part (ii) of Property \ref{P:instreward} and the fact that
$\pi^1_t P^{N-n} \leq_{st} \pi^1_t P^{N-n+1}$ (Property \ref{P:preserveorder}).
\\\\
(i) When $O_t(N)= 1$ (i.e. $n = N$), assume $O_t(N-1)=N$ without loss of generality. Then $A_{Nm}O_t(N)= O_t(N-1)=N$.
Therefore,
\begin{align}
&V_t(A_{Nm}O_t,\pi^{1:N}_t) -V_t(O_t,\pi^{1:N}_t) \nonumber\\
=& (\pi^{N}_t-\pi^1_t)R 
+ \beta \sum_{i < L} \pi^{N}_t(i) V_{t+1}(S(A_{Nm}O_t),\pi^{1:N-1}_tP,P_i)
 +\beta \sum_{i \geq L} \pi^{N}_t(i) V_{t+1}(A_{Nm}O_t,\pi^{1:N-1}_tP,P_i)\nonumber\\
& - \beta \sum_{i < L} \pi^1_t(i)V_{t+1}(SO_t,P_i,\pi^{2:N}_tP)
  - \beta \sum_{i \geq L} \pi^1_t(i) V_{t+1}(O_t,P_i,\pi^{2:N}_tP)
\nonumber\\
=& (\pi^{N}_t-\pi^1_t)R 
+\beta \sum_{i < L} \pi^{N}_t(i)
[V_{t+1}(S(A_{Nm}O_t),\pi^{1:N-1}_tP,P_i)-  V_{t+1}(A_{Nm}O_t,\pi^{1:N-1}_tP,P_i)] 
+\beta V_{t+1}(A_{Nm}O_t,\pi^{1:N}_tP)\nonumber\\
& - \beta \sum_{i < L} \pi^1_t(i)V_{t+1}(SO_t,P_i,\pi^{2:N}_tP)
  - \beta \sum_{i \geq L} \pi^1_t(i) V_{t+1}(O_t,P_i,\pi^{2:N}_tP)\nonumber\\
=& (\pi^{N}_t-\pi^1_t)R 
+ \beta\sum_{i < L} \pi^{N}_t(i)
[V_{t+1}(S(A_{Nm}O_t),\pi^{1:N-1}_tP,P_i)-  V_{t+1}(A_{Nm}O_t,\pi^{1:N-1}_tP,P_i)] 
\nonumber\\
& +\beta\sum_{i < L} \pi^1_t(i)[V_{t+1}(A_{Nm}O_t,P_i,\pi^{2:N}_tP)-V_{t+1}(SO_t,P_i,\pi^{2:N}_tP)]\nonumber\\
&  +\beta \sum_{i \geq L} \pi^1_t(i)[V_{t+1}(A_{Nm}O_t,P_i,\pi^{2:N}_tP)- V_{t+1}(O_t,P_i,\pi^{2:N}_tP)]\nonumber\\
\leq& (\pi^{N}_t-\pi^1_t)R + \beta \sum_{i < L} \pi^{N}_t(i)
[V_{t+1}(S(A_{Nm}O_t),\pi^{1:N-1}_tP,P_i)-  V_{t+1}(A_{Nm}O_t,\pi^{1:N-1}_tP,P_i)] \nonumber\\
\leq& (\pi^{N}_t-\pi^1_t)R + \beta \sum_{i < L} \pi^{N}_t(i)(h-P_iR).
\label{pfP:funlift:eq2}
\end{align}
The three equalities in (\ref{pfP:funlift:eq2}) follow from the recursive equation and the linearity of the function $V_{t+1}$ (
(\ref{eqVt}) and Lemma \ref{lm:funlin}).
The last inequality in (\ref{pfP:funlift:eq2}) follows from the induction hypothesis for Property \ref{P:funlift}, the fact that $S(A_{Nm}O_t) = A_{N1}(A_{Nm}O_t)$, and $A_{Nm}O_t(N)=O_t(N-1)=N$ and the fact that $P_i \leq_{st} P_iP$ for $i<L$ by Property \ref{P:PMForder}.
\\
The first inequality in (\ref{pfP:funlift:eq2}) is true because of the following:\\
For $i< L$, $P_i\leq_{st}P_{L-1}\leq_{st}\pi^l_tP $ for all $l$ by Property \ref{P:PMForder}.
Then,
\begin{align}
&V_{t+1}(A_{Nm}O_t,P_i,\pi^{2:N}_tP)-V_{t+1}(SO_t,P_i,\pi^{2:N}_tP) \nonumber\\
=&V_{t+1}(W_{m(m-1)}...W_{32}W_{21}SO_t,P_i,\pi^{2:N}_tP)-V_{t+1}(SO_t,P_i,\pi^{2:N}_tP) 
\leq 0.
\label{pfP:funlift:eqA1}
\end{align}
The equality in (\ref{pfP:funlift:eqA1}) is true because $A_{Nm}O_t=W_{m(m-1)}...W_{32}W_{21}SO_t$.
The inequality in (\ref{pfP:funlift:eqA1}) follows by repeatedly using Property \ref{P:funswitch2} at $t+1$ and the fact that for $i< L$, $P_i\leq_{st}\pi^l_tP$ for all $l$.
\\
For $i\geq L$, $P_i\geq_{st}P_{L}\geq_{st}\pi^l_tP $ for all $l$ by Property \ref{P:PMForder}.
Then,
\begin{align}
&V_{t+1}(A_{Nm}O_t,P_i,\pi^{2:N}_tP)-V_{t+1}(O_t,P_i,\pi^{2:N}_tP) \nonumber\\
=&V_{t+1}(W_{m(m+1)}...W_{(N-1)(N-2)}W_{N(N-1)}O_t,P_i,\pi^{2:N}_tP)-V_{t+1}(O_t,P_i,\pi^{2:N}_tP) 
\leq 0.
\label{pfP:funlift:eqA2}
\end{align}
The equality in (\ref{pfP:funlift:eqA2}) is true because $A_{Nm}O_t=W_{m(m+1)}...W_{(N-1)(N-2)}W_{N(N-1)}O_t$.
The inequality in (\ref{pfP:funlift:eqA2}) follows by repeatedly using Property \ref{P:funswitch2} at $t+1$ and the fact that for $i\geq L$, $P_i\geq_{st}\pi^l_tP$ for all $l$.
\\\\
Let $v$ be the vector such that 
\begin{align}
&v_i = \left\lbrace \begin{array}{rr}
	  R_i+\beta(h- P_iR), & \text{for } i<L\\
	  R_i,            & \text{for } i\geq L  
\end{array}\right. .
\end{align}
For $i\geq L$ we have
\begin{align}
v_{i+1}-v_i = R_{i+1}-R_i \geq 0.
\end{align}
For $i= L-1$, 
\begin{align}
v_{L}-v_{L-1} = R_{L}-R_{L-1}-\beta (h-P_{L-1}R) \geq 0;
\label{pfP:funlift:eqLL1}
\end{align}
the inequality if (\ref{pfP:funlift:eqLL1}) holds because of condition (A\ref{A:reward}).
\\
For $i< L-1$, we have
\begin{align}
v_{i+1}-v_i  =& R_{i+1}-R_i -\beta(P_{i+1}-P_i)R \nonumber\\
            \geq & R_{i+1}-R_i -\beta(P_{i+1}-P_i)M \nonumber\\
            \geq &0.
            \label{pfP:funlift:eqv}
\end{align}
The first inequality in (\ref{pfP:funlift:eqv}) follows from part (ii) of Property \ref{P:instreward};
the last inequality in (\ref{pfP:funlift:eqv}) follows from condition (A\ref{A:reward}).
\\
Consequently, $v_i$ increases with $i$. Then, from part (i) of Property \ref{P:instreward} and the fact that 
$\pi^{N}_t\leq_{st}P_K $ we obtain
\begin{align}
&V_t(A_{Nm}O_t,\pi^{1:N}_t) -V_t(O_t,\pi^{1:N}_t) \nonumber\\
\leq & (\pi^{N}_t-\pi^1_t)R + \beta \sum_{i < L} \pi^{N}_t(i)(h-P_iR)\nonumber\\
= & \pi^{N}_tv-\pi^1_tR \nonumber\\
\leq & P_Kv-\pi^1_tR \nonumber\\
=& h-\pi^1_tR
\label{pfP:funlift:eqlast}
\end{align}
The first inequality in (\ref{pfP:funlift:eqlast}) follows from (\ref{pfP:funlift:eq2}).
The second inequality in (\ref{pfP:funlift:eqlast}) follows from part (i) of Property \ref{P:instreward}, the fact that $v_i$ increases with $i$, and the fact that $\pi^{N}_t\leq_{st}P_K $.
The last equality in (\ref{pfP:funlift:eqlast}) follows from the observation that
\begin{align}
P_K v = P_KR+\beta \sum_{i<L}p_{Ki}(h-P_iR)=h.
\end{align}
This completes the proof of the induction step for Property \ref{P:funlift} at $t$,
and the proof of the entire induction step.
\end{myproof}

\section*{Acknowledgment}
This work was supported in part by National Science Foundation (NSF) Grant CCF-1111061 and NASA grant NNX12A0546.

\ifCLASSOPTIONcaptionsoff
  \newpage
\fi



%
%

\bibliographystyle{IEEEtran}
\bibliography{nchannelref}
%

%






\end{document}